\documentclass[lettersize,journal]{IEEEtran}
\usepackage{amsmath,amsfonts}
\usepackage{algorithmic}
\usepackage{algorithm}
\usepackage{array}
\usepackage{capt-of}
\usepackage[caption=false,font=normalsize,labelfont=sf,textfont=sf]{subfig}
\usepackage{textcomp}
\usepackage{stfloats}
\usepackage{color,soul}
\usepackage{url}
\usepackage{verbatim}
\usepackage{graphicx}
\usepackage{cite}
\usepackage{pifont}
\usepackage{multirow}
\usepackage{mathtools}
\usepackage{dsfont}
\NewDocumentCommand{\set}{o m}{%
  \IfNoValueTF{#1}
    {\{#2\}}
    {\{#1 \mid #2\}}%
}
\usepackage{makecell}
\usepackage{booktabs}
\usepackage{kotex}
\usepackage{xcolor}
\usepackage{fancyhdr,lipsum} 
\usepackage{hyperref}
\newcommand{\github}[1]{%
   \href{#1}{\faGithubSquare}%
}
\DeclarePairedDelimiter\abs{\lvert}{\rvert}%

\begin{document}

\title{PR-CARA: Proactive V2X Resource Allocation with Extended 1-Stage SCI and Deep Learning-based Sensing Matrix Estimator}

\author{Taesik Nam, \IEEEmembership{Student Member, IEEE}, Seungjae Lee, Kiwoong Park, Sunbeom Kwon, Nathan Jeong, \IEEEmembership{Senior Member, IEEE}, Han-Shin Jo, \IEEEmembership{Member, IEEE}, and Jong-Gwan Yook, \IEEEmembership{Senior Member, IEEE}

\thanks{This work was supported in part by Institute of Information \& communications Technology Planning \& Evaluation (IITP) grant funded by the Korea Government (MSIT) (No. 2022-0-01053), and in part by the National Research Foundation of Korea (NRF) grant funded by the Korea government (MSIT) (No. RS-2024-00346319)(Corresponding authors: Han-Shin Jo; Jong-Gwan Yook)}

\thanks{Taesik Nam and Jong-Gwan Yook are with the Department of Electrical and Electronic Engineering, Yonsei University, Seoul, 03722 South Korea (e-mail: ts.nam@yonsei.ac.kr; jgyook@yonsei.ac.kr)}

\thanks{Han-Shin Jo, Seungjae Lee, Kiwoong Park, and Sunbeom Kwon are with the Department of Automotive Engineering, Hanyang University, Seoul, 04763 South Korea (e-mail: hsjo23@hanyang.ac.kr; freitag97@hanyang.ac.kr; akdclgnd@hanyang.ac.kr; ksb33174@hanyang.ac.kr)}

\thanks{Nathan Jeong is with the Department of Electrical and Computer Engineering, The University of Alabama, Tuscaloosa, AL 35487, USA (email:
shjeong@eng.ua.edu)}}

\markboth{Journal of \LaTeX\ Class Files,~Vol.~0, No.~0, None~2024}%
{Shell \MakeLowercase{\textit{et al.}}: A Sample Article Using IEEEtran.cls for IEEE Journals}
\maketitle

\begin{abstract}
Distributed resource allocation algorithms differ from centralized methods by relying on locally collected information for resource selection, leading to a low vehicle-to-everything (V2X) communication quality of service (QoS) in high-traffic congestion. To overcome these challenges, this study proposes a proactive received signal strength indicator (RSSI)-based collision avoidance resource allocation (PR-CARA) algorithm. This algorithm features an extended 1-stage SCI system, which is a critical component that enables resource monitoring of adjacent vehicle user equipment (VUE). Monitored resources were then processed through  a deep learning-based proactive RSSI estimator. The estimated proactive RSSI helps avoid resource selection, which leads to packet collisions, thereby significantly reducing the occurrence of this issue during resource allocation. The proposed algorithm is tested in a cooperative adaptive cruise control (CACC)-based platoon driving scenario that requires ultra-reliable and low-latency communication (URLLC) performance. Simulation results demonstrate that the proposed deep-learning-based proactive resource allocation algorithm, with the extended 1-stage SCI system, reduces packet collisions and improves the transmission signal-to-interference-plus-noise ratio (SINR), thereby significantly enhancing communication reliability compared to the benchmark resource allocation algorithm.
\end{abstract}

\begin{IEEEkeywords}
Wireless communication, Intra-platoon cooperation, resource allocation scheme, packet collision avoidance, V2X networks, CACC-based platoon driving.
\end{IEEEkeywords}

\section{Introduction}
Cooperative-intelligent transport systems (C-ITS) aim to automate the operation and management of transportation systems by sharing and utilizing advanced transportation technologies and information, such as electronic control and communication for transportation modes and facilities. This technology can reduce traffic accidents and improve traffic efficiency, and research is actively being conducted in this area, as shown in \cite{citsTITS1, citsTITS2, citsTITS3, citsTITS4}.

Autonomous driving technology for C-ITS begins with an early-stage basic safety message service and evolves toward the ultimate goal of fully autonomous driving. Vehicle-to-everything (V2X) communication is a core technology in this progression, with different quality of service (QoS) requirements for each application. For example, services such as Local Dynamic Map (LDM) and entertainment applications require QoS with high data rates. For example, services such as LDM and entertainment applications require QoS with high data rates. This is supported through vehicle-to-infrastructure (V2I) or vehicle-to-network (V2N) communication, which supports the transmission and reception of large-scale data. In contrast, applications that ensure vehicle safety require low-latency, high-reliability communication QoS, which is achieved via a vehicle-to-vehicle (V2V) link that minimizes latency through direct communication. The 3rd Generation Partnership Project (3GPP) carried out standardization efforts to meet these technical requirements. 3GPP Releases 14 and 15 focused on the standardization of Long Term Evolution V2X (LTE-V2X)\cite{cv2x}, with Release 16 extending these efforts to introduce 5G-based New Radio V2X (NR-V2X) communication technology\cite{nrv2x}. In addition, to fulfill the high QoS requirements of V2X applications, centralized resource allocation methods such as Mode 3 of LTE-V2X and Mode 1 of NR-V2X, along with distributed resource allocation algorithms like Mode 4 of LTE-V2X and Mode 2 of NR-V2X, have been standardized. This study focuses on V2V communication using a distributed resource allocation method, specifically targeting low-latency applications.

Distributed resource allocation allows each vehicle to independently determine its radio resources. In this method, each vehicle measures the received signal strength indicator (RSSI) of radio resources divided into the time and frequency domains. Each vehicle randomly selects and transmits a radio resource from a candidate single subframe resource (CSR) list consisting of resources with low RSSI or those that are not reserved. However, distributed resource allocation algorithms based on semi-persistent scheduling (SPS) suffer from degraded reliability owing to packet collisions and decoding errors that occur for the following reasons. The first is the sudden interference that occurs during the resource retention period, and the second is the degradation in the accuracy of RSSI detection owing to hidden/exposed terminal issues arising during the resource selection process. This problem worsens in environments with heavy communication traffic, such as during rush hours \cite{wilabV2Xsim}. Against this background, two major research categories have emerged to improve resource allocation algorithms.

The first is to improve the accuracy of the detected RSSI. Distributed resource allocation algorithms exclude resources from the CSR list whose average RSSI exceeds a certain threshold. However, vehicle networks experience high mobility and rapid channel variability, which degrade the accuracy of the detected RSSI. In response to this, studies \cite{extendedDataYang, extendedDataZhang, extendedDataAli1, extendedDataAli2,extendedDataAli3, extendedDataSabeeh1, extendedDataSabeeh2, extendedDataSabeeh3, extendedDataWang1,extendedDataWang2} have aimed to address the inaccuracy of the detected RSSI by incorporating additional information into the packets transmitted by each vehicle.

Ali et al. \cite{extendedDataAli1, extendedDataAli2, extendedDataAli3} proposed a system that adds the currently used resource information to the Physical Sidelink Shared Channel (PSSCH) Transport Block (TB), which includes the data payload and 2-stage Sidelink Control Information (SCI), shared with other vehicles. This information is used in resource allocation algorithms to prevent packet collisions and improve communication reliability. However, this requires the prerequisite that the TB be successfully decoded, which is difficult to achieve in high-channel congestion environments. To address this limitation, Sabeeh et al. and Wang et al. \cite{extendedDataSabeeh1, extendedDataSabeeh2, extendedDataSabeeh3, extendedDataWang1, extendedDataWang2} proposed solutions that utilize the extra bits of the reserved field in the 1-stage SCI, transmitted via the Physical Sidelink Control Channel (PSCCH). In addition, Wang et al. \cite{extendedDataWang1, extendedDataWang2} proposed a collaborative resource allocation algorithm that uses these extra bits to share information, including the presence of platoon vehicles, location of adjacent vehicles, resource possession, and direction of vehicle movement. These studies attempted to indirectly solve the problem of inaccurately detected RSSI by sharing additional information (resources and vehicle mobility). Consequently, communication reliability was improved; however, the fundamental issue of detected RSSI inaccuracy remains unresolved.

The second approach attempted to improve the resource selection process. In distributed resource allocation algorithms, vehicles randomly select radio resources from the CSR and maintain the selected radio resources for a certain period. To address the packet collision problem during resource maintenance, Bazzi et al. \cite{selectProcedureBazzi} proposed an algorithm to determine when to reallocate resources while maintaining them. In addition, Molina et al. \cite{selectProcedureMolina} proposed an algorithm to reallocate radio resources based on the mobility information of surrounding vehicles. What these studies have in common is their focus on minimizing the packet collision period by adjusting the resource maintenance period or preventing collisions by avoiding radio resources with a high probability of collision. However, they rely on additional information transmitted by other vehicles and are limited by the difficulty in receiving reliable information from other vehicles in highly congested environments.

Recently, studies \cite{ccncYang, drlYang, jychoi, aiBasedSelectSang, aiBasedSelectReshma, drlBasedSelectLiang, drlBasedSelectJu, drlBasedSelectLee, drlBasedSelectFarzanullah} using AI-based algorithms, such as deep learning and reinforcement learning, have been published to improve the resource selection procedure mentioned above. Sang et al. \cite{aiBasedSelectSang} proposed an algorithm that estimates and allocates the optimal power based on the received CSI using a deep learning model. More recently, a deep reinforcement learning (DRL)-based approach was employed by Liang et al. \cite{drlBasedSelectLiang} to solve the spectrum-sharing problem using a multi-agent deep Q-learning algorithm. Building on this, Ju et al. \cite{drlBasedSelectJu} designed a multi-agent DRL resource allocation model that minimizes system processing delays while addressing the security issues of load-balancing algorithms. Farzanullah et al. \cite{drlBasedSelectFarzanullah} proposed a DRL model to maximize V2V and V2I packet transmission rates within the target delay time. In this model, each agent is classified into three classes and performs platoon leader selection, user association, and power control for V2I links, as well as channel and power-level allocation for V2V link support. The proposed algorithm is highly valuable because it demonstrates superior performance compared to the benchmark algorithms presented in this paper and targets actual V2X applications.

RL-based algorithms enable agents to make optimal decisions based only on the local observations of the operating environment. This makes them highly suitable for communication applications and preferable over supervised learning-based deep learning approaches. However, this RL-based approach faces challenges such as model convergence issues and performance degradation due to environmental changes. Reinforcement learning algorithms learn from their interactions with the environment; however, vehicular communication environments are highly variable. Moreover, in V2X communication, resource allocation is a complex problem that must account for various dimensions such as frequency and time, which results in a large state-action space. This degrades the convergence of reinforcement learning models and is deepened in multi-agent DRL models. To address this challenge, \cite{drlBasedSelectFarzanullah} attempted to solve the convergence issue by separating agents according to the dimensions in which they were designed and by creating state-action spaces of appropriate size for each agent. 

This study builds on and extends the learning model simplification proposed by \cite{drlBasedSelectFarzanullah} and additional information exchange systems introduced by Ali et al. and Sabeeh et al. \cite{extendedDataAli1, extendedDataAli2, extendedDataAli3, extendedDataSabeeh1, extendedDataSabeeh2, extendedDataSabeeh3}. To achieve this, this study focused on the fundamental principles of distributed resource allocation algorithms. In distributed resource allocation algorithms, the degradation of communication QoS is primarily caused by the inaccuracy of the detected RSSI, which fails to account for the effects of hidden and exposed nodes. Thus, enhancing the accuracy of the detected RSSI can improve communication QoS. This reasoning provides insight that communication QoS can be improved by estimating the physical impact of subsequent access nodes (hidden nodes) and revoked nodes (exposed nodes) on the minimum resource allocation unit, the physical resource block (PRB).

This insight suggests that the learning scope can be reduced from the entire V2X network to a single PRB, which has not been previously attempted. Furthermore, this can be realized through the sharing of additional information (subsequent transmission resources), implemented via the extended 1-stage SCI proposed in this study. In addition, by reducing the learning scope to a single PRB, the training objective can be simplified to estimate the proactive RSSI, which reflects the interference effects of the hidden and exposed nodes using the detected RSSI and additional information. In particular, the interference effects of the hidden and exposed nodes follow a standardized channel model, which enables the use of a physics-based AI methodology \cite{physicsBasedAI, tsnamTIFS} for model training. By generating data based on physical phenomena, this approach can effectively reflect the variations in data caused by changes in traffic congestion (vehicle density). This ensures data diversity and enhances model robustness, representing a key contribution of this study. Thus, this approach resolves the challenges of model convergence and performance degradation due to environmental changes experienced by traditional RL-based approaches.

In summary, the research presented in this study makes the following main contributions:
\begin{itemize}
    \item Improvement in detected RSSI accuracy: The proposed system enhances inter-vehicle resource monitoring and facilitates the estimation of proactive RSSI, which accounts for the effects of hidden and exposed nodes. This approach improves RSSI accuracy, unlike previous studies, and we demonstrate that it can significantly enhance QoS.
    \item Simplifying the optimization problem: Distributed resource allocation can be framed as an optimization problem aimed at maximizing URLLC performance. However, the inherent complexity of this problem often makes it challenging to determine an optimal solution. Proactive RSSI simplifies the optimization problem, enabling the identification of an effective resource allocation.
    \item Securing model robustness by utilizing physics-based AI: This study narrows the learning scope by focusing on the core principles of resource allocation, thereby enabling the application of physics-based AI. Consequently, the model effectively reflects data variations induced by changes in vehicle density, leading to improved robustness.   
    \item Verification across diverse environments and services: The proposed resource allocation algorithm exhibits stable performance across various scenarios. Improved QoS results are demonstrated in vehicular communication environments characterized by low congestion, high congestion, and real-world event scenarios. Additionally, the verification is conducted in a semi-realistic V2X network environment by integrating the Simulation of Urban Mobility (SUMO) traffic simulator \cite{SUMO} with the WiLabV2Xsim \cite{wilabV2Xsim} communication simulator.
\end{itemize}
This paper aimed at strengthening these contributions, is organized as follows: Section \ref{sec.systemModel} describes the model used to verify the proposed algorithm. Section \ref{sec.proposedMethod} provides a detailed description of the proposed deep-learning-based proactive resource allocation algorithm. Section \ref{sec.simulationResult} presents the simulation results and evaluates the communication performance using the packet delivery ratio ($PDR$) and inter-packet gap ($IPG$) results, which are directly related to reliability and latency. Finally, Section \ref{sec.conclusion} summarizes the results, reconfirms the aforementioned contributions, and concludes the paper.

\section{System Model}\label{sec.systemModel}
This section describes the system model used to verify the performance of the proposed algorithm. The proposed algorithm is validated in a cooperative adaptive cruise control (CACC)-based platoon driving scenario, in which periodic and aperiodic message transmission services are operated. For this purpose, a CACC-based platoon driving scenario and periodic/aperiodic message transmission network system model are described, and a resource allocation problem is formulated.

\subsection{CACC-based Platoon Driving Service}
The explosive increase in vehicle traffic during rush hour degrades traffic efficiency and increases the risk of accidents. An adaptive cruise control (ACC)-based platoon driving \cite{accRef1, accRef2, accRef3, accRef4} service was proposed to address these issues and improve vehicle traffic efficiency. The ACC system estimates the dynamics of the front vehicle (e.g., speed, acceleration, and inter-vehicle gap) using sensors, such as radar, lidar, camera, and ultrasonic waves, and maintains the distance between vehicles. ACC focuses on providing reliable and comfortable driving services for each vehicle. However, because of the inevitable limitations of sensors, errors occur when estimating the distance between vehicles. These errors destabilize ACC-based control and consequently reduce driving stability.

\begin{figure}[tb!]
  \centering
  \includegraphics[clip, width = 0.8\columnwidth]{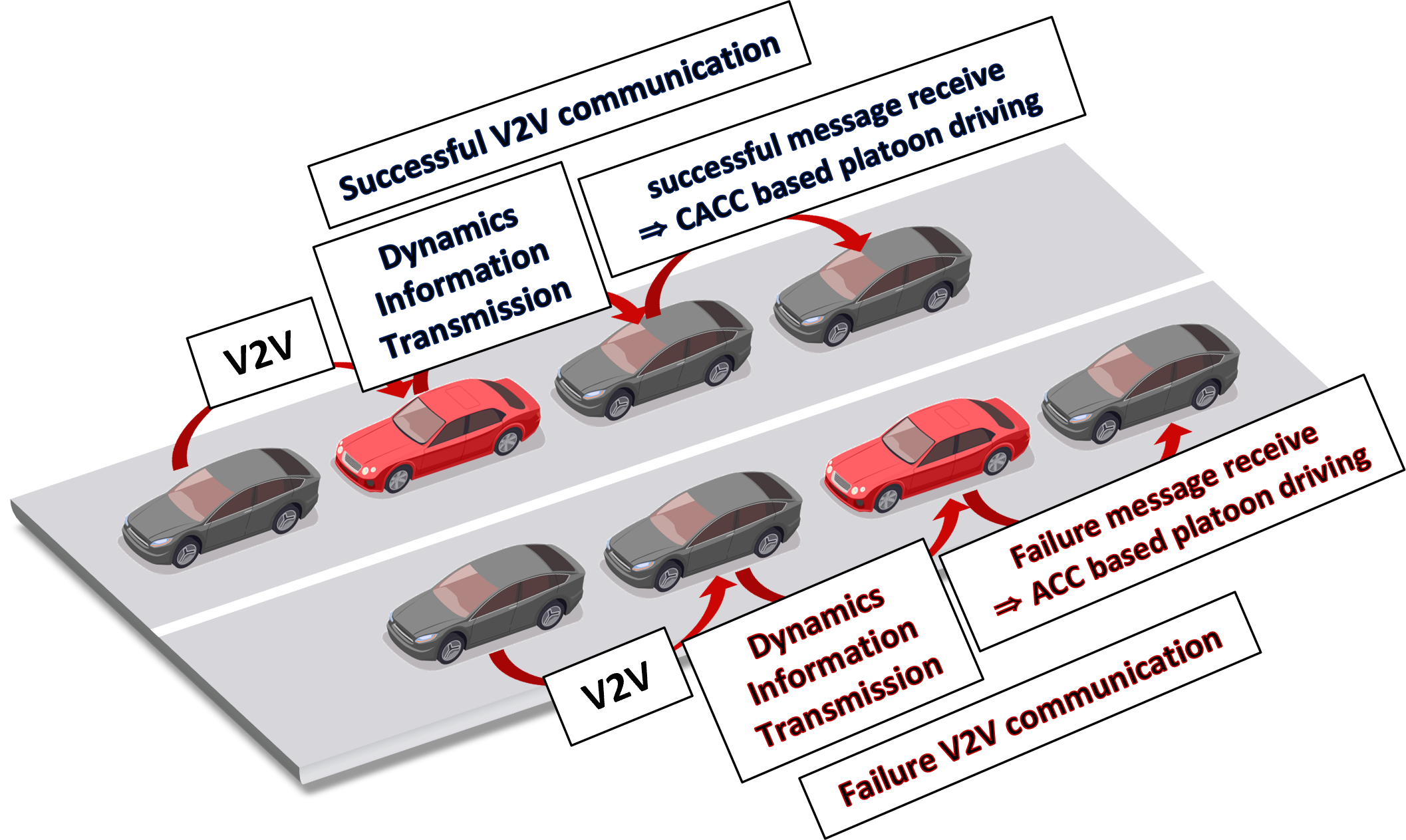}
  \caption{CACC-based driving service scenario}
  \label{fig.caccBasedDriving}
\end{figure}

\begin{figure}[tb!]
  \centering
  \includegraphics[clip, width = 0.8\columnwidth]{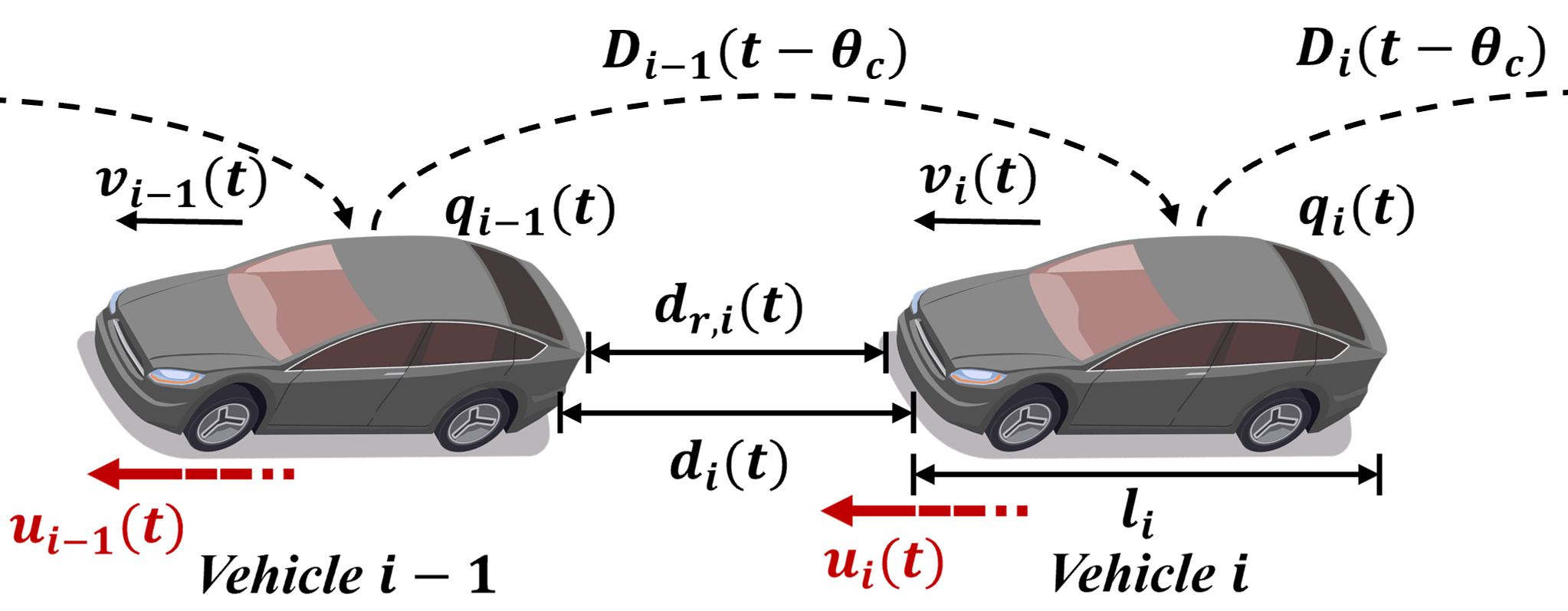}
  \caption{Dynamics parameter of the CACC system}
  \label{fig.dynamicsParamOfCacc}
\end{figure}

To overcome these technical limitations, the CACC system was proposed \cite{caccETSI, caccTR, platoonTR, caccPractical, caccTITS2, ccNEDO, caccTITS3, caccTITS4}. A CACC is a system in which each vehicle is connected through a V2V link, as shown in Fig. \ref{fig.caccBasedDriving}, and dynamics information is shared between platoon vehicles through the V2V link. The primary dynamics information of the CACC system is shown in Fig. \ref{fig.dynamicsParamOfCacc}, where $v_i$, $q_i$, and $u_i$ represent the velocity, position, and acceleration of vehicle $i$, respectively, and $l_i$ represents the total length of vehicle $i$. where $D_{i-1}(t-\theta_c)$ represents the dynamics information (e.g., $v_i$, $q_i$, and $u_i$) with a communication delay of $\theta_c$, in which vehicle $i-1$ transmits to vehicle $i$. CACC systems can ensure faster responses and shorter inter-vehicle distances than sensor-dependent ACC systems. This speed-harmonization service is a technology that increases road capacity, smoothens traffic flow, and ensures traffic safety. However, CACC’s services cannot be maintained unconditionally. As shown in Fig. \ref{fig.caccBasedDriving}, a short-IPG and high-reliability V2V link can ensure stable dynamics information transmission and provide a stable driving experience through the CACC system. However, in a long IPG and low-reliability V2V link, dynamics information transmission becomes unstable, and CACC services conditionally convert into ACC services \cite{caccPractical}. Consequently, the CACC system poses a challenging issue from the perspective of V2V communication because of its demand for a short Transmission Interval (TTI) ($\theta_c$) and high-reliability performance \cite{platoonTR, ccNEDO}.

\begin{figure*}[ht!]
  \centering
  \includegraphics[width = 0.8\textwidth]{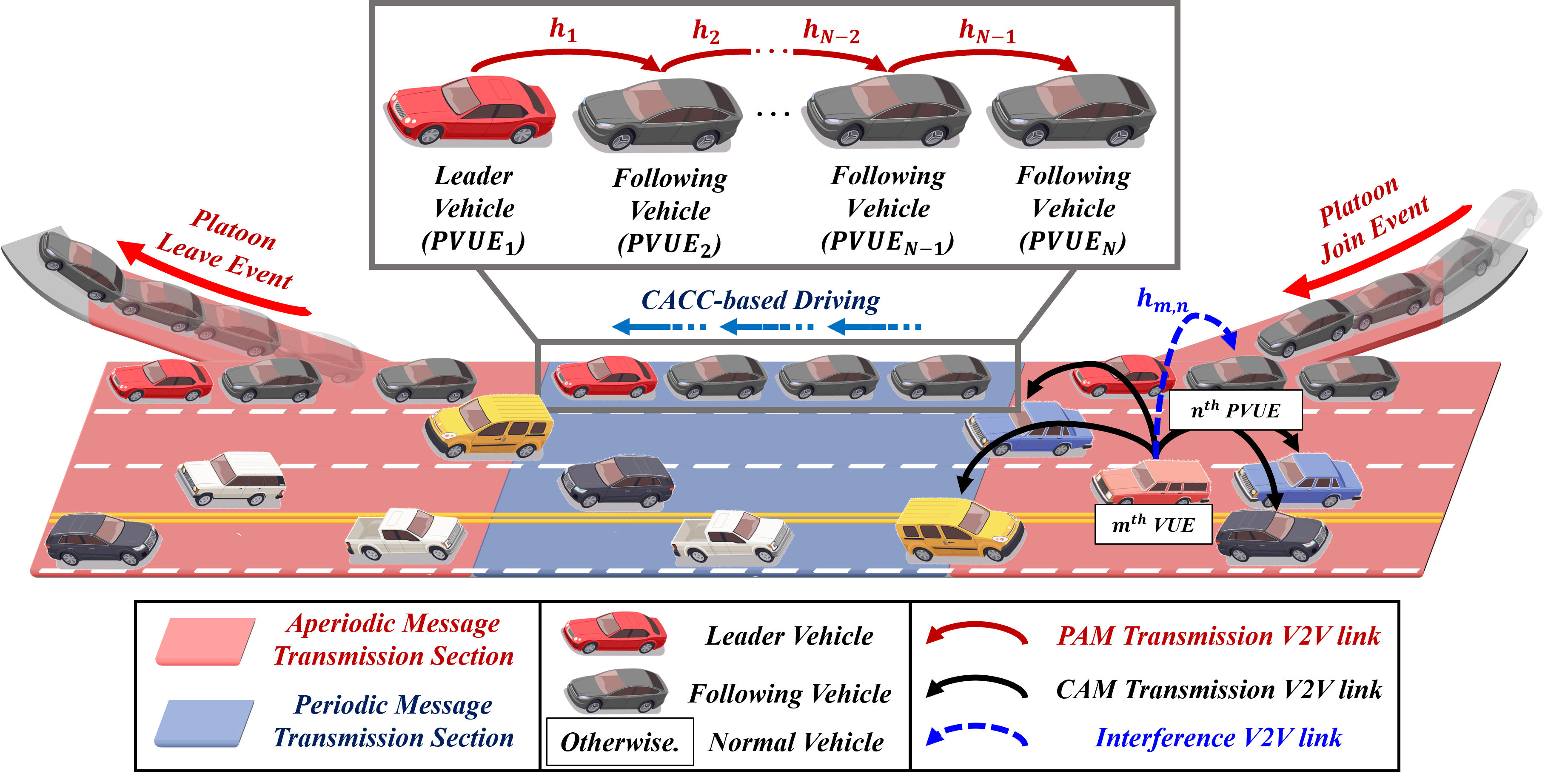}
  \caption{A system model for verification of the proposed algorithm: The red road section operates an event-triggered service (aperiodic), and the blue road section operates a CACC-based driving service (periodic). The red and black solid arrows represent the PAM and CAM transmission V2V links, respectively, while the blue dotted arrows indicate the interference that the CAM V2V link causes in the PAM V2V link.}
  \label{fig.highwaySystemModel}
\end{figure*}

\begin{figure}[tb!]
  \centering
  \includegraphics[clip, width = 0.8\columnwidth]{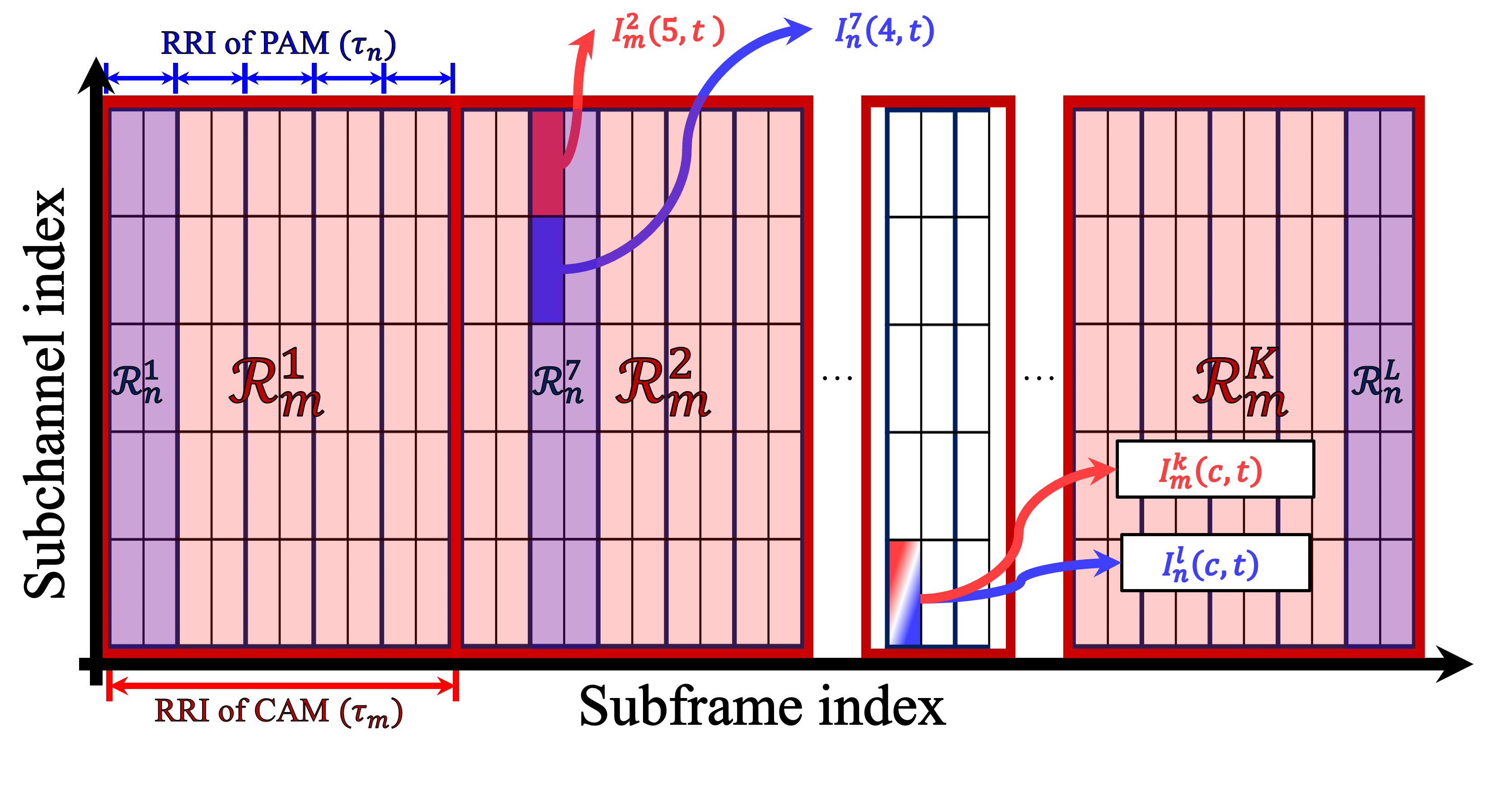}
  \caption{Wireless network resource pool of CACC-based platoon driving system model}
  \label{fig.resourcePoolV2X}
\end{figure}

\subsection{Road Scenario and Network Model}
This study assumes the system model shown in Fig. \ref{fig.highwaySystemModel}. The environment consists of $M$ vehicle user equipment (VUE), which broadcasts cooperative awareness messages (CAM), and $N$ platoon vehicle user equipment (PVUE), which broadcasts unicast platoon cooperative awareness messages (PAM). VUEs are regular passenger cars, and the location of each VUE is uniformly distributed according to the density $\rho$ (vehicles/km). The main service target of platooning is the transportation industry, and PVUEs are accordingly distributed in the rightmost lane. The distance between VUEs varied owing to changes in the driving velocity. In contrast, PVUEs operate based on a CACC-based platoon driving service; therefore, the distance between vehicles in the same group remains constant.

The proposed algorithm is verified for two road sections. The first is CACC-based driving on a highway section, and the second is platoon joining/leaving a merge/branch road section. The highway section has a network configuration that shares periodic dynamics information with CACC-based driving. In the merge/branch road section, the original platooning vehicles operate using CACC-based driving, whereas the platoon head vehicle (HV) and platoon joining/leaving vehicles (JV/LV) transmit and receive event-triggered aperiodic messages according to a defined service flow \cite{joiningLeaving}. In this study, we aim to emphasize the robustness of the proposed algorithm by analyzing the communication QoS in two road sections.

The sets of vehicle indices transmitting the CAM and PAM packets are represented as $\mathcal{M}$ and $\mathcal{N}$, respectively, where $\mathcal{M} = \left \{ 1, 2, ..., M \right \}$ and $\mathcal{N} = \left \{ 1, 2, ..., N \right \}$. In the system model, communication links for two types of messages are considered: a CAM-broadcast-V2V-link for periodic safety message transmission and a PAM-unicast-V2V-link for platooning driving support (CACC, joining/leaving). The CAM-broadcast-V2V-link of the $m$-th VUE does not have a specific destination network node. Instead, each vehicle decodes the CAM received from other vehicles for periodic safety message transmission services. Conversely, the PAM-unicast-V2V-link of the $n$-th PVUE has a designated destination network node according to the platooning support service.

The resource pool of the V2X communication system is defined as a set of resources $(\mathcal{R})$ divided into the time and frequency domains. The resource pool comprises multiple subframes (LTE-V2X) or slots (NR-V2X) in the time domain. The subframe index set defined during the simulation period in this study is $\mathcal{T}$, where $\mathcal{T} = \left \{1, 2, ..., T\right \}$. In the frequency domain, the resource pool consists of subchannels comprising multiple resource blocks (RBs). The number of RBs used for message transmission varies depending on the defined message size and  modulation coding scheme (MCS) used; accordingly, more than one subchannel may be used. The subchannel index set is $\mathcal{C}$, where $\mathcal{C} = \left \{1, 2, ..., C \right \}$. Finally, the resource set $\mathcal{R}$ is represented as $\mathcal{R}= \left \{I(c, t)\;|\;c\in \mathcal{C}, t\in \mathcal{T} \right \}$, which is used to define the resource allocation problems. Fig. \ref{fig.resourcePoolV2X} illustrates the resource pool, assuming it consists of 5 subchannels and that both CAM and PAM use subchannels for transmission. In addition, the VUE and PVUE had different resource reservation interval (RRI). The figure depicts a situation where the VUE's RRI is 100 ms, and the PVUE's RRI is 20 ms. 

Resource set $\mathcal{R}$ can be expressed as a union of resource subsets divided according to the RRI, as shown in (\ref{eq.csrSubset}):
\begin{equation}
    \begin{aligned}
    \mathcal{R} & = \bigcup_{k\in \mathcal{K}}\mathcal{R}^k_m \\
    & = \bigcup_{l\in \mathcal{L}}\mathcal{R}^l_n,
    \end{aligned}
\label{eq.csrSubset}
\end{equation}
where the subsets $\mathcal{R}^k_m$ and $\mathcal{R}^l_n$ represent the resource sets used when the $m$th VUE and $n$th PVUE transmit the $k$th CAM and $l$th PAM, respectively. $\mathcal{K} = \left \{1, 2, ..., K \right \}$ and $\mathcal{L} = \left \{1, 2, ..., L \right \}$ are defined, where $K$ and $L$ represent the number of CAM and PAM transmissions, respectively, during the simulation period. Consequently, resource subsets $\mathcal{R}^k_m$ and $\mathcal{R}^l_n$ are defined as follows:
\begin{equation}
\mathcal{R}^k_m = \left\{
  I^k_m(c, t) \;\middle|\;
  \begin{aligned}
  & c\in \mathcal{C}, t\in \mathcal{T} \\
  & (k-1)\cdot \tau_{m} < t \leq k\cdot \tau_{m}
  \end{aligned}
\right\},
\label{eq.camSubset}
\end{equation}
\begin{equation}
\mathcal{R}^l_n = \left\{
  I^l_n(c, t) \;\middle|\;
  \begin{aligned}
  & c\in \mathcal{C}, t\in \mathcal{T} \\
  & (l-1)\cdot \tau_{n} < t \leq l\cdot \tau_{n}
  \end{aligned}
\right\},
\label{eq.pamSubset}
\end{equation}
where $\tau_m$ and $\tau_n$ are the $RRI$ values of the CAM and PAM, respectively, in milliseconds. Where $I^k_m(c, t)$ and $I^l_n(c, t)$ have integer values of $0$ for resources that are not allocated for message transmission and $1$ for resources that are allocated for message transmission and must meet the following conditions:
\begin{equation}
    \sum_{I^k_m(c, t)\in\mathcal{R}^k_m} I^k_m(c, t) = 1,
    \label{eq.camResourceIndicator}
\end{equation}
\begin{equation}
    \sum_{I^l_n(c, t)\in\mathcal{R}^l_n} I^l_n(c, t) = 1.
    \label{eq.pamResourceIndicator}
\end{equation}
This means that for each $k$ and $l$, one resource must be allocated from the defined sets of $\mathcal{R}^k_m$ and $\mathcal{R}^l_n$, and no other resources should be allocated. Thus, the definitions above imply that the $m$-th VUE and $n$-th PVUE transmit the $k$-th CAM and $l$-th PAM, respectively, by selecting a resource from the resource subsets $\mathcal{R}^k_m$ and $\mathcal{R}^l_n$ for each transmission. This is used to define the signal-to-interference-plus-noise ratio (SINR) of the received message.

The channel power gain of $m$-th VUE and $n$-th PVUE links is defined as
\begin{equation}
    h^k_{m, n} = g_{m, n}\xi_{m, n}AL^{-\kappa}_{m, n}\overset{\underset{\mathrm{\Delta}}{}}{=}g_{m, n}\alpha_{m, n},
\label{eq.camChannelPowerGain}
\end{equation}
\begin{equation}
    h^l_n = g_n\xi_nAL^{-\kappa}_n\overset{\underset{\mathrm{\Delta}}{}}{=}g_n\alpha_{n},
\label{eq.pamChannelPowerGain}
\end{equation}
where $\alpha_n = \xi_nAL^{-\kappa}_n$ is a large-scale slow fading channel parameter, $\xi_n$ is a log-normal shadow fading random variable with standard deviation $\varrho$, $A$ is a pathloss constant, $L_n$ is the distance between the $n$-th PVUE and the communication target, $\kappa$ is the decay exponent, and $g_n$ is the small-scale fast fading channel parameter. The described content pertains to the $n$-th PVUE link channel power gain in (\ref{eq.pamChannelPowerGain}); however, the parameters in  (\ref{eq.camChannelPowerGain}) have the same physical meaning. However, because this study focuses on the performance of the PAM, the channel power gain of the CAM only considers the interference link, and $L_{m, n}$ refers to the distance between the $m$-th VUE and the $n$-th PVUE link communication targets.

In summary, the SINR of the $l$-th PAM transmission by the $n$-th PVUE vehicle is defined as follows:
\begin{equation}
    \gamma^l_n = \frac{P_{n}h^l_nI^l_n(c, t)}{\sigma^2 + \sum_{m\in\mathcal{M}}P_{m}h^k_{m, n}I^k_m(c, t)},
    \label{eq.caccV2vLinkSinr}
\end{equation}
where $P_m$ and $P_n$ are the transmission powers of the $m$-th VUE and $n$-th PVUE, respectively, and $\sigma$ is the noise power. $c$ and $t$ represent the subchannel and subframe indices, respectively, of the resources used by the $n$-th PVUE link. When the $m$-th VUE link uses the same $c$ and $t$ values, that is, when $I^k_m(c, t) = 1$, it is considered as interference. The SINR of the $n$-th PVUE link, defined in this manner, is subsequently discussed in the communication QoS analysis process and problem formulation.

\subsection{Sensing-based Resource Allocation Scheme}
The CACC-based platoon driving service is a V2X application that requires low latency and employs a distributed resource allocation method. The distributed execution resource allocation algorithm primarily operates through standardized sensing-based semi-persistent scheduling (SB-SPS) \cite{sbsps}.

The SB-SPS is divided into two stages. The first stage is the RSSI sensing stage, and the second stage is the resource selection stage. In the sensing stage, each vehicle senses the RSSI in a minimum resource allocation unit, the PRB. The RSSI sensing duration for the subsequent resource selection stage is 1000 ms for periodic services and 100 ms for aperiodic services.

\begin{figure}[htb!]
\centering
\subfloat[]{
\includegraphics[clip, width = 0.8\columnwidth]{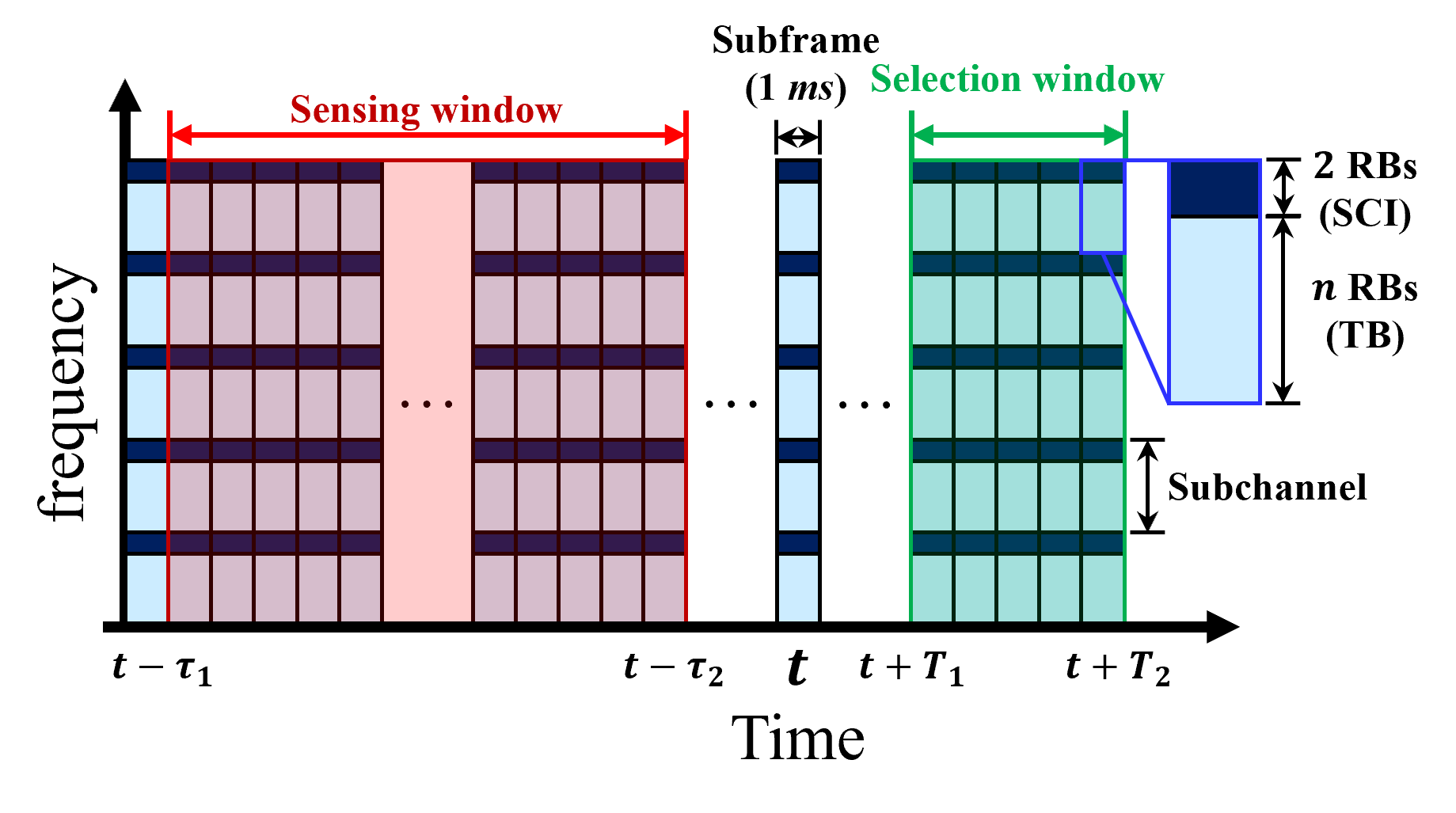}
\label{fig.sbsps(a)}
}

\subfloat[]{
\includegraphics[clip, width = 0.8\columnwidth]{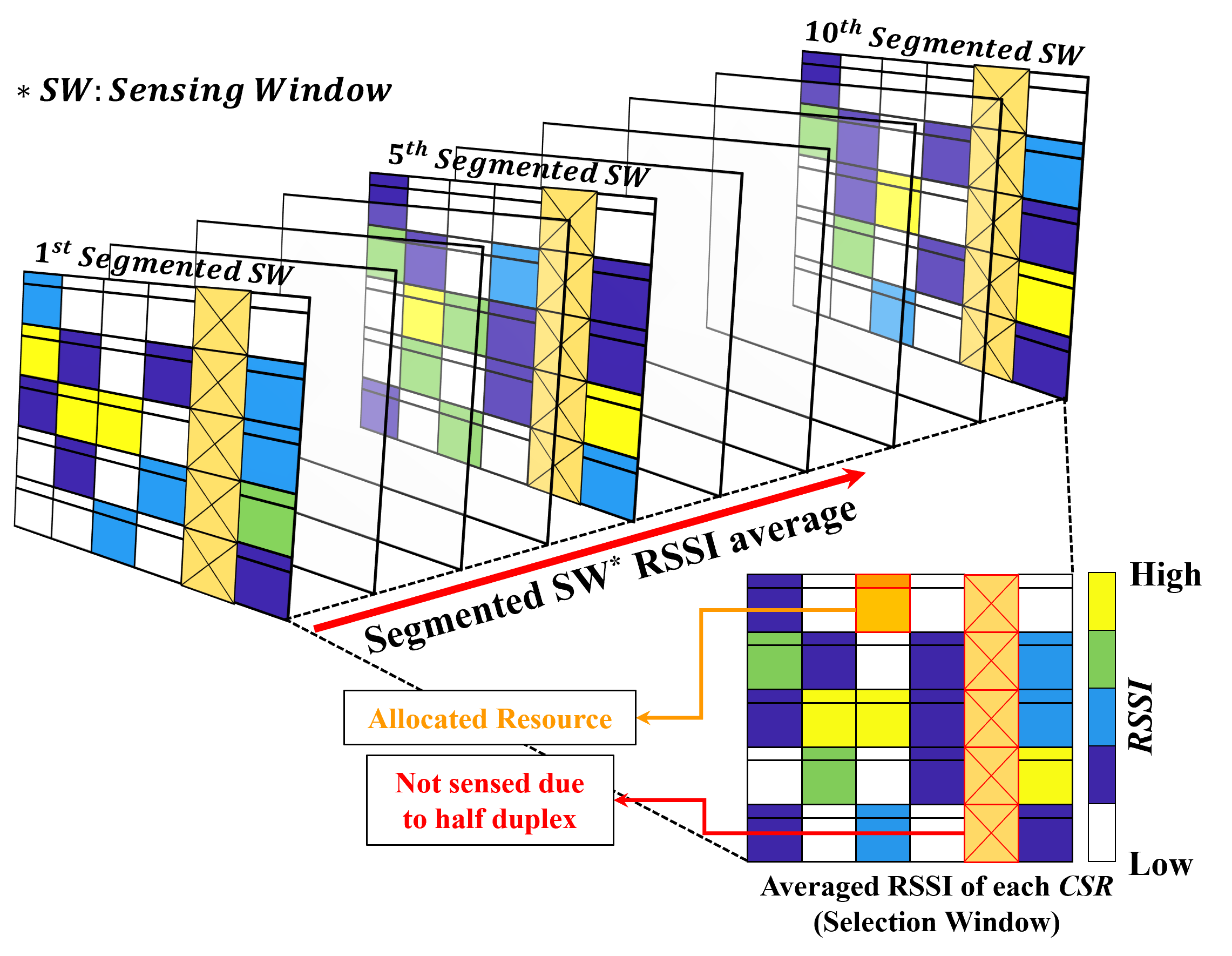}
\label{fig.sbsps(b)}
}
\caption{(a) Configuration of sensing and selection windows for SB-SPS; (b) Average RSSI calculation process for each $CSR$ for SB-SPS}
\label{fig.sbsps}
\end{figure}

In the resource selection stage, the resources in the selection window are chosen based on the RSSI values measured during the sensing window. The selection stage consisted of the following four steps:
\begin{itemize}
    \item Step 1: The $m$-th VUE or $n$-th PVUE divides the sensing window based on the RRI as defined in Fig. \ref{fig.sbsps}-(a). This segmented sensing window is then used to calculate the average RSSI value measured for each CSR, as shown in Fig. \ref{fig.sbsps}-(b). Resources with an average RSSI value exceeding the reference signal received power (RSRP) threshold are stored in the CSR set ($CSR^k_m$ or $CSR^l_n$). For simplicity in the following explanation, descriptions will be given from the perspective of the $m$-th VUE; however, the radio resources for the $n$-th PVUE are determined in the same manner.
    \item Step 2: If $\abs{CSR^k_m} \geq 0.2\abs{\mathcal{R}^k_m}$, proceed to Step 3. Otherwise, increase the RSRP threshold by 3 dB and repeat Step 1. The notation $\abs{\cdot}$ represents the total number of elements in a set. Thus, $\abs{\mathcal{R}^k_m}$ represents the total elements of the VUE's resource subset $(\mathcal{R}^k_m)$ in the selection window.
    \item Step 3: VUE stores the resource to $CSR^k_m$ that has an RSSI value in the lower 20\% and has received a reserved field of PSCCH with 0.
    \item Step 4: The VUE randomly selects one resource from $CSR^k_m$ and allocates the selected resource for transmitting the next packet. In addition, it sets the reselection counter (RC) value for the SPS configuration.
\end{itemize}

The above process and Fig. \ref{fig.sbsps} describe the resource allocation process through SB-SPS. However, the direct application of SB-SPS to V2X applications requiring low-latency and high-reliability services is limited for two reasons. The first is the variation in the received SINR over time. The SB-SPS uses the SPS scheme to maintain the selected resources for a certain period according to the configured RC. These resources were selected based on the previously measured RSSI of the sensing window (Fig. \ref{fig.sbsps}), but this cannot reflect the unexpected interference from packet transmission nodes that occur during the resource reservation. This affects the reliability of the V2X services.

The second issue is packet collisions and decoding errors caused by hidden or exposed transmission nodes from adjacent VUEs. This occurs when assuming that the $m$-th VUE selects resources at $t_m$, and the adjacent VUE allocates resources in the interval $[t_m - \tau_m, t_m]$. This is not reflected in the RSSI of the sensing window of the $m$-th VUE. Hidden nodes increase the correlation between the CSR lists of the $m$-th VUE and its adjacent VUEs, which increases the possibility of occupying the same subframe resources and exposed nodes cause the $m$-th VUE to exclude PRBs with a low RSSI when forming its CSR list. This causes packet collisions due to HD errors and packet errors due to decoding failure, which degrades communication reliability.

To solve the first issue, we applied a sensing-based dynamic scheduling (SB-DS) method that prevents unexpected interference and guarantees a high received SINR for each transmission. However, this method alone cannot completely resolve the problems caused by hidden/exposed node transmissions. The solution to this problem is further discussed in Section \ref{sec.proposedMethod}.

\subsection{Problem Formulation}
To define an optimization problem that maximizes the PAM packet transmission and reception rates, it is essential to describe the PVUE packet transmission process. PVUE transmits a packet at $c$ and $t$ that satisfies $I_n^l(c, t) = 1$ among the elements of $\mathcal{R}_n^l$ defined in the previous subsection. The transmitted packet is received using the SINR calculated using (\ref{eq.caccV2vLinkSinr}). For the packet to be decoded, the $\gamma_n^l$ value must be greater than $\gamma^l_{n, 0}$, where $\gamma^l_{n, 0}$ is the minimum SINR required for packet decoding. This value varies depending on the payload and MCS of the packet. The reception or transmission errors of these packets can be formalized as follows:
\begin{equation}
\mathds{1}_{\left \{\gamma^l_n \geq  \gamma^l_{n, 0}\right \}}(c, t) = \begin{cases}
\begin{aligned}
 1 \quad & \text{if} \quad \gamma_{n}^l(c, t) \geq \gamma^l_{n, 0}\\
 0 \quad & otherwise.
\end{aligned}
\end{cases}
\end{equation}
Ultimately, the optimization problem for maximizing the packet reception rate can be summarized by selecting the optimal $c$ and $t$ across all PVUEs $(\forall n \in \mathcal{N})$ and each transmission $(\forall l \in \mathcal{L})$. The formula is as follows:
\begin{equation}
    \begin{aligned}
        \underset{\left \{ c,\;t\right \}}{\text{maximize}} \quad & \frac{\sum_{n\in\mathcal{N}} \sum_{l\in\mathcal{L}}\mathds{1}_{\left \{\gamma^l_n \geq  \gamma^l_{n, 0}\right \}}(c,\;t)}{\sum_{n\in\mathcal{N}} \sum_{l\in\mathcal{L}}I_n^l(c,\;t)},\\
        \text{s.t.} \quad & I_n^l(c,\;t) = 1, \\
        & I_n^l(c,\;t) \in \mathcal{R}^l_n, \\
        & \sum_{I_n^l(c,\;t)\in \mathcal{R}^k_m} I_n^l(c,\;t) = 1, \\
        & c \in \mathcal{C},\; t \in\mathcal{T}. \\
    \end{aligned}\label{eq.optimizationProblem}
\end{equation}
The denominator of the objective function represents the total number of packets transmitted by all PVUEs during the simulation time, whereas the numerator represents the number of successfully decoded packets. Thus, the defined objective function corresponds to the PDR. Additionally, the constraints represent the resource allocation rules that must be followed within the system model, as explained in (\ref{eq.csrSubset})–(\ref{eq.pamResourceIndicator}).

The defined optimization problem is extremely complex because it must be considered in various dimensions, such as time (subframe) and frequency(subchannel), and determining the optimal solution is challenging. SB-SPS, a standardized resource allocation methodology, selects $\left \{c,\;t\right \}$ based on the detected RSSI and reserved indicator received via the PSCCH. However, SB-SPS does not completely solve the unexpected interference caused by hidden/exposed node transmissions. Section \ref{sec.proposedMethod} addresses how to solve hidden/exposed node issues and simplifies the complex optimization problems of distributed resource allocation algorithms.

\begin{figure}[t!]
  \centering
  \includegraphics[clip, width = 0.8\columnwidth]{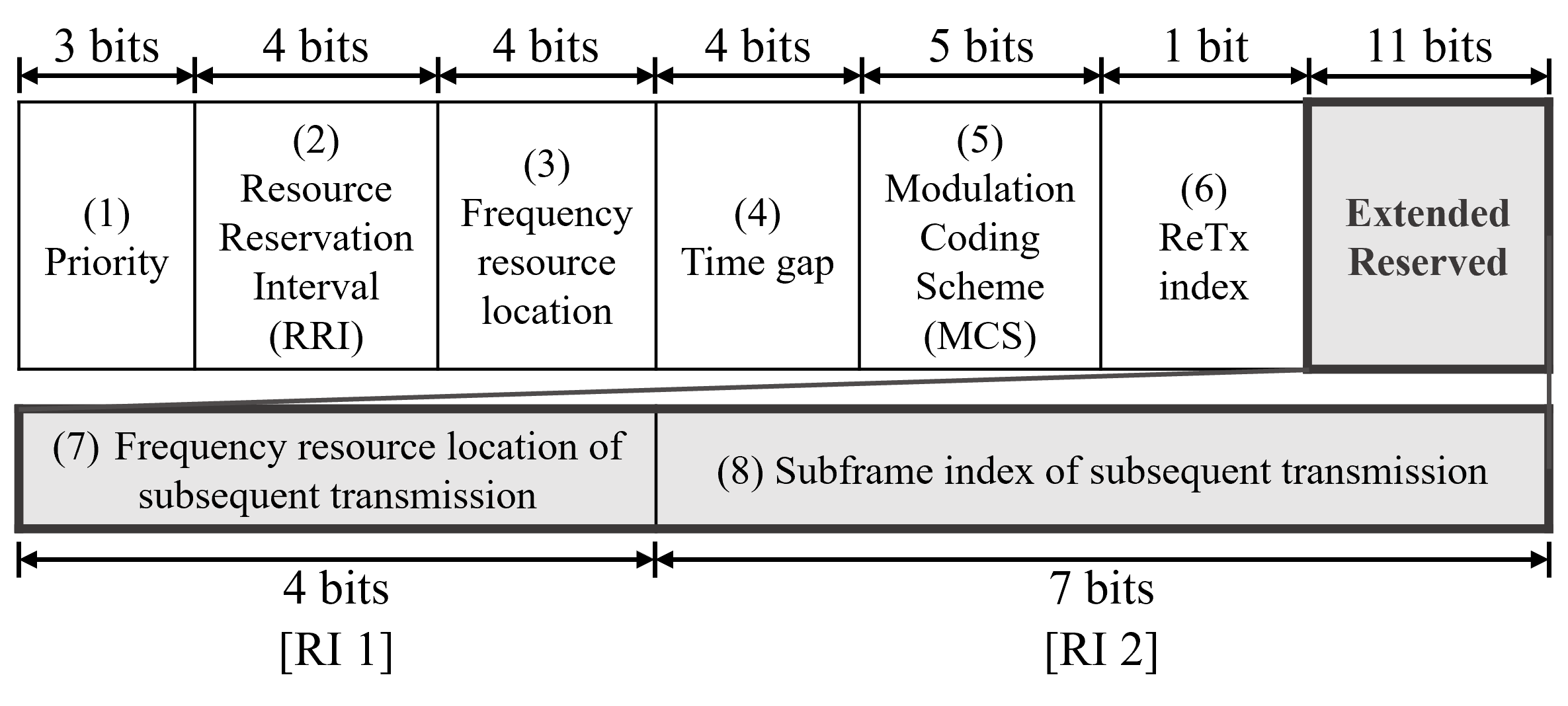}
  \caption{Extended 1-stage SCI message format}
  \label{fig.extendedSCI}
\end{figure}

\section{Proactive Resource Allocation Scheme}\label{sec.proposedMethod}
This section describes the proposed system and methodology for addressing the major issues caused by hidden/exposed node transmissions. The proposed system uses a standardized V2X communication system that enables information acquisition from hidden and exposed transmission nodes. The acquired information is used to obtain a proactive RSSI that reflects the interference effects of hidden/exposed nodes using a deep learning-based estimator. In addition, this approach translates the defined resource allocation optimization problem into a straightforward minimum RSSI detection problem, thereby enabling quasi-optimal resource selection.

\subsection{Extended 1-stage SCI for Proactive Resource Allocation}
The proposed algorithm is based on the resource monitoring between adjacent vehicles. However, resource monitoring is restricted in distributed resource allocation methodology. To this end, we propose a Resource Information (RI) exchange system using SCI transmitted over PSCCH. SCI is transmitted in two stages, and the 1-stage SCI transmitted to  PSCCH provides information for decoding the transport block (TB). SCI uses a low-order modulation scheme for stable transmission and requires low SINR for message decoding. The proposed algorithm leveraged these stable transmission characteristics. To this end, a 1-stage SCI message format structure for RI exchange, as shown in Figure \ref{fig.extendedSCI}, is proposed. This structure is an extension of the ”reserved“ field of the original 1-stage SCI and has the same bit length of 32 bits as in the original message format. PVUE and VUE transmit subsequent transmission RI utilizing the fields (RI1 and RI2) in the extended 1-stage SCI. Consequently, the proposed system enables resource monitoring between adjacent vehicles.

The (RI1, RI2) set in SCI in PVUE $l$-th transmission is set as follows:
\begin{itemize}
\item RI1: Refers to frequency resource location information used in $(l+1)$-th transmission, and the value is calculated and assigned in the same manner as in (3) of Fig. \ref{fig.extendedSCI}.
\item RI2: Refers to the offset $(t_n^{l+1} - t_n^{l})$ between the scheduled $(l+1)$-th transmission subframe index and the current $l$-th transmission subframe index. At this time, the offset value has an integer value between [1, 127], is encoded and then assigned. For example, if the offset value is $\eta_n$, it implies that the $(l+1)$-th transmission is scheduled in the $(t_n^{l} + \eta_n)$-th subframe index.
\end{itemize}
If the decoded (RI1, RI2) is $0$, the resource is used for subsequent transmissions. Conversely, these resources are not used for subsequent transmissions if they are not $0$. This replaced the role of the "reserved" field in the original 1-stage SCI. Thus, the extended 1-Stage SCI enables monitoring of the subsequent transmission of resource information between adjacent vehicles.

\subsection{Proactive RSSI Estimator Design via Deep-Neural Network Model}
The extended 1-Stage SCI allows each vehicle to monitor the RI of the adjacent VUE and PVUE. However, the proposed RI exchange system alone cannot address the fundamental issue of RSSI inaccuracy. Because distributed resource allocation relies on the detected RSSI, this issue is a primary factor in performance degradation. In this subsection, we propose a deep-learning model for proactive RSSI estimation that reflects the interference effects of hidden/exposed transmission nodes.

The extended 1-stage SCI enables resource monitoring between adjacent vehicles. Additionally, the CAM that the vehicle periodically broadcasts contains GNSS-based position information\cite{camRef1, camRef2, camRef3}. In other words, the position and RI of the adjacent vehicles (VUE, PVUE) can be obtained from the received packets, and this information is used to estimate a proactive RSSI that reflects the interference effect of hidden/exposed nodes. In this study, the design of the proactive RSSI estimator used a deep learning-based model to be robust against the effects of channel noise.

Returning to the main topic, this subsection describes the design process of a deep-learning-based proactive sensing matrix estimator. The design process is divided into four stages: first, acquiring training data; second, designing the neural network model; third, defining the error and objective functions for model training; and fourth, tuning hyperparameters for the optimization algorithm.

The first step is to establish strategies for acquiring training data for the deep learning models. The design of a deep learning-based model is based on data-driven modeling. In other words, the performance of the trained model is determined based on data quality. Consequently, strategies for acquiring data and ensuring data diversity are the most important aspects of deep learning-based model design. The adopted data acquisition and training strategy were based on a physics-based AI methodology \cite{physicsBasedAI}. The physics-based AI methodology formalizes and models physical phenomena in a wireless channel and network access environment to acquire data and train models based on them. The V2X environment is characterized by high mobility and requires the consideration of various vehicle densities. In other words, this is a challenging environment for data collection and diversity. To solve this problem, this study adopts a physics-based AI strategy that leverages the physical phenomena in a V2X environment (e.g., wireless channel and network access congestion levels) to acquire data and train the model.

\begin{figure}[t!]
  \centering
  \includegraphics[clip, width = 0.8\columnwidth]{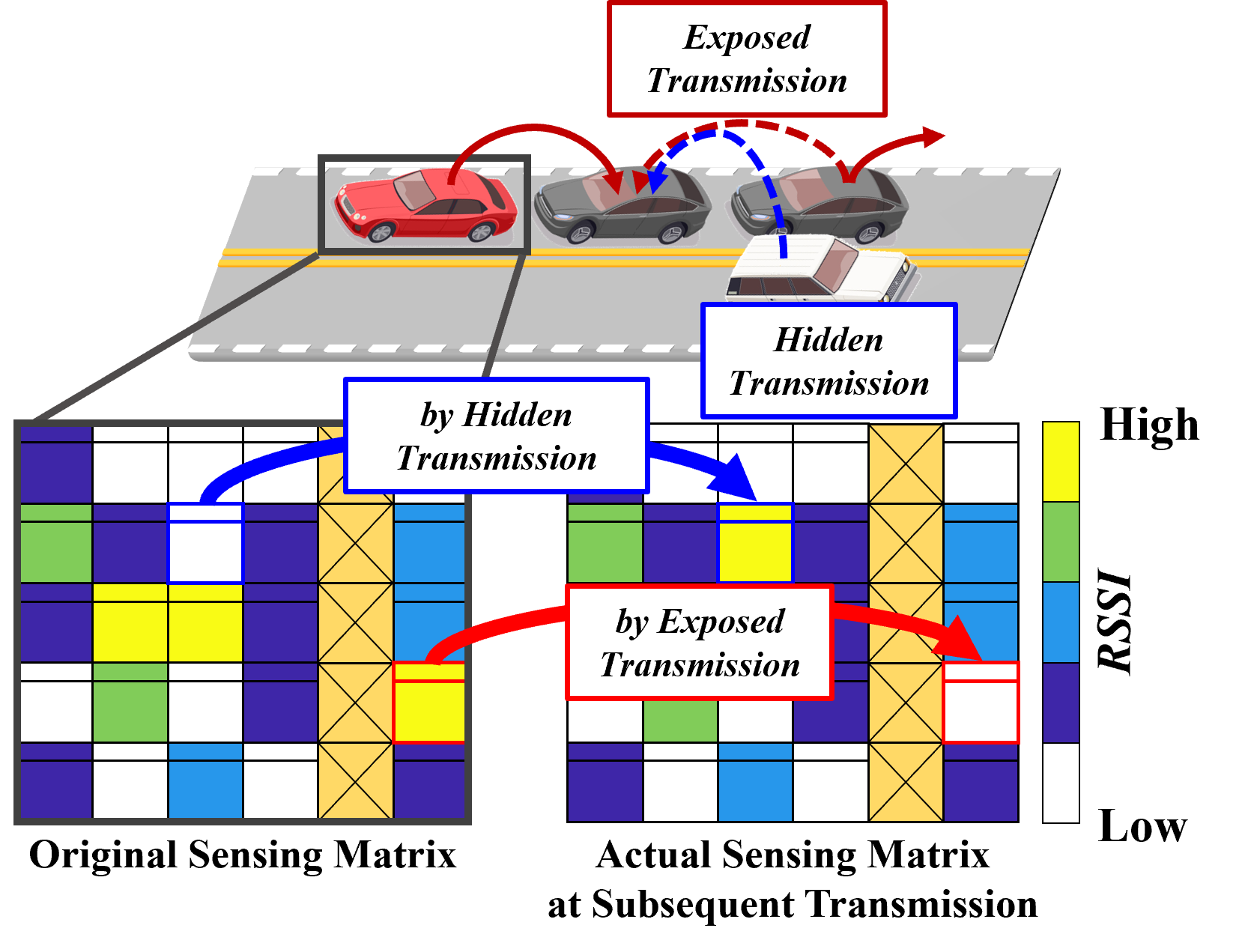}
  \caption{The original sensing matrix does not reflect the effects of hidden/exposed transmissions; the actual sensing matrix at subsequent transmissions and the original sensing matrix show a difference.}
  \label{fig.hiddenAndRemovedInterference}
\end{figure}

The design of the proactive RSSI estimator aims to account for unexpected interference from hidden or exposed transmissions in the sensing matrix. This interference is not reflected in the original RSSI of the sensing matrix, as shown in Fig. \ref{fig.hiddenAndRemovedInterference}. Against this background, a data-acquisition strategy using physics-based AI considers three factors: first, the distance and number of original transmission vehicles (original interference); second, the distance and number of hidden transmission vehicles (hidden interference); and third, the distance and number of exposed transmission vehicles (exposed interference).

The first factor determines the original RSSI, whereas the second and third factors determine the hidden and exposed RSSI. Consequently, the relationship between the original RSSI and proactive RSSI is formulated as follows:
\begin{equation}
    \begin{aligned}
        \underbrace{\tilde{\varepsilon}^P_{n, l}(c,\;t)}_{\text{Proactive RSSI}} \quad = \quad & \underbrace{\tilde{\varepsilon}^{O}_{n, l}(c,\;t)}_{\text{original RSSI}} \\
        + \quad & \underbrace{\tilde{\varepsilon}^{H}_{n, l}(c,\;t)}_{\text{Hidden RSSI}} \\
        - \quad & \underbrace{\tilde{\varepsilon}^{E}_{n, l}(c,\;t)}_{\text{Exposed RSSI}},
    \end{aligned}
    \label{eq.proactiveSensingMatrix}
\end{equation}
where the original RSSI is designed for the first factor, with the distance range set to a random value between [150, 1000] meters and the number of vehicles set to a random integer value between [0, 6]. The hidden/exposed RSSI is designed for the second/third factor with the distance range set to a random value between [3, 150] m and the number of vehicles set to a random integer value between [0, 1]. (\ref{eq.proactiveSensingMatrix}) is applied to create an environment that mimicked the physical phenomena of V2X communication, including wireless channel and network access congestion. In other words, $\tilde{\varepsilon}$ represents the data generated by physics-based AI methodology. This environment, designed using the above parameters, reflects the radio resource status across low-to-high congestion levels. This mimicking environment supplied data for training the proactive RSSI estimator, reflecting the effects of hidden/exposed transmission nodes across various V2X congestion levels.

The second step in the design of the deep learning model focused on the neural network structure used for the proactive RSSI estimator. This estimator has a multilayer perceptron (MLP) structure consisting of five input dimensions, one output dimension, and five hidden layers with 64 nodes each and uses a Rectified Linear Unit (ReLU) activation function. Recently, various neural networks, such as convolutional neural network (CNN)-based and recurrent neural network (RNN)-based structures, have been adopted in deep-learning research. However, the estimator to be designed does not require the advantages of CNN’s spatial feature extraction or RNN’s sequential data feature extraction. Therefore, we adopted the MLP structure, which is simple and advantageous for analyzing the relationship between input and output data.

The third step in the deep learning model design describes the objective function (loss function) for model training. The designed loss function is based on (\ref{eq.proactiveSensingMatrix}). The loss function $L(\theta)$ is set to mean square error (MSE) between the proactive RSSI (label) and the estimated RSSI (model output), which is expressed as
\begin{equation}
    L(\theta)=\mathbb{E}[(\;\underbrace{\tilde{\varepsilon}^P_{n, l}(c, t)}_{\text{Proactive RSSI}}-\underbrace{\mathcal{F}(\tilde{\varepsilon}^O_{n, l}(c, t), I^H, D^H, I^E, D^E ; \theta)}_{\text{Estimated RSSI}}\;)^2],
    \label{eq.lossFunctionForEstimator}
\end{equation}
where $\tilde{\varepsilon}^O_{n, l}(c, t)$ denotes the original RSSI. $I^{H}(c, t)$ and $I^{E}(c, t)$ are information identified through the extended 1-stage SCI, indicating the presence or absence of hidden/exposed transmission at resource $(c, t)$. $D^{H}(c, t)$ and $D^{E}(c, t)$ are identified through the TB containing GNSS-based position information, which indicates the distance from the hidden/exposed transmission node at resource $(c, t)$. In conclusion, the proactive RSSI estimator is trained to minimize (\ref{eq.lossFunctionForEstimator}).

The fourth step in the deep learning model design is to tune the hyperparameters of the optimization algorithm for estimator training. Adaptive moment estimation (Adam) is selected for the optimization. Adam is a deep learning optimization algorithm combining Momentum and RMSProp's advantages. The main hyperparameters for the Adam optimizer are the learning rate $(\alpha)$, exponential moving average $(\beta_1)$ of Momentum, and exponential moving average $(\beta_2)$ of RMSProp. To train the model, $(\alpha)$ is set to 0.05, $(\beta_1)$ as 0.9, and $(\beta_2)$ as 0.999.

\begin{figure}[tb!]
  \centering
  \includegraphics[clip, width = 0.8\columnwidth]{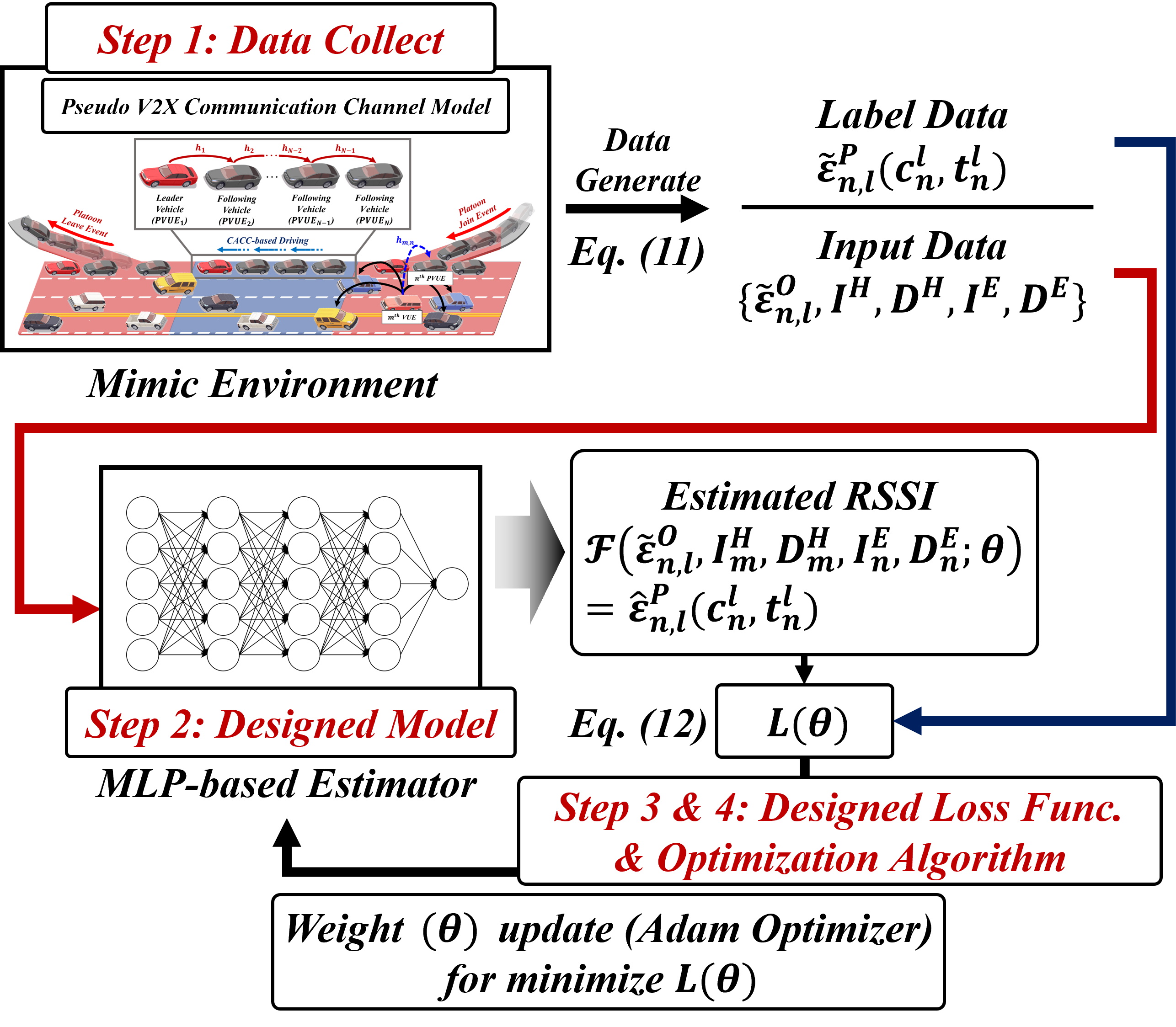}
  \caption{Summarized proactive sensing matrix estimator design process}
  \label{fig.TrainingProcessForProactiveSensingMatrixEstimator}
\end{figure}

The four-step process for designing the proactive sensing matrix estimator described earlier is summarized in Fig. \ref{fig.TrainingProcessForProactiveSensingMatrixEstimator}.

\subsection{Distributed Resource Allocation Execution used Proactive RSSI}
This subsection describes the utilization of a trained estimator for the V2X vehicle communication resource allocation algorithm. The proactive RSSI estimator transforms (\ref{eq.optimizationProblem}) into the following solvable problem: The transformed optimization problem is expressed as
\begin{equation}
    \begin{aligned}
        \underset{\left \{ c, t\right \}}{\text{minimize}} \quad & \hat{\varepsilon}^P_{n, l}(c, t) \\   
        \text{s.t.} \quad & I_n^l(c,\;t) = 1, \\
        & I_n^l(c,\;t) \in \mathcal{R}^l_n, \\
        & \sum_{I_n^l(c,\;t)\in \mathcal{R}^k_m} I_n^l(c,\;t) = 1, \\
        & c \in \mathcal{C},\; t \in\mathcal{T}. \\
    \end{aligned}
    \label{eq.transformedOptimizationProblem}
\end{equation}
In the original optimization problem (\ref{eq.optimizationProblem}), the effects of the hidden/exposed nodes are not reflected in the optimization problem, which prevents the determination of the optimal solution. Proactive RSSI provides information to avoid packet collisions and decoding errors based on the subsequent transmission information of adjacent vehicles. Consequently, the proactive RSSI enables the original problem to be transformed into an optimization problem that allows us to find a quasi-optimal solution. \ref{eq.transformedOptimizationProblem}) represents the minimum RSSI detection problem, where the quasi-optimal solution is to select the $\left \{c, t\right \}$ pair with the minimum proactive RSSI $\hat{\varepsilon}^P_{n,l}$ from the resource set $\mathcal{R}_n^l$. This quasi-optimal solution avoids packet collisions and maximizes SINR $\gamma^l_n$ in (\ref{eq.caccV2vLinkSinr}). Consequently, it provides a simple and practical solution that enhances the PDR and objective function of (\ref{eq.optimizationProblem}).

\begin{figure}[htb!]
  \centering
  \includegraphics[clip, width = 0.8\columnwidth]{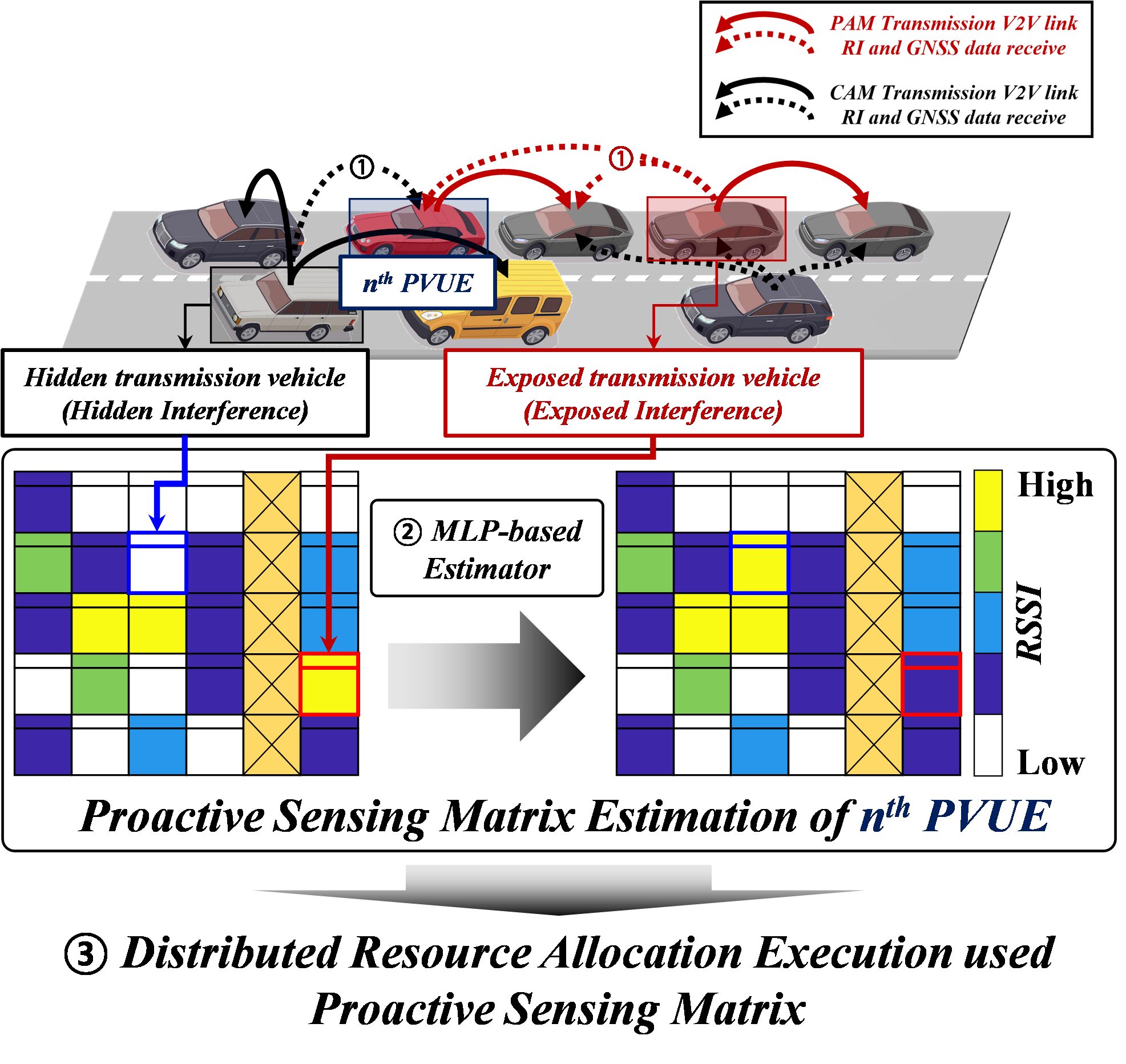}
  \caption{Distributed resource allocation execution process used designed proactive sensing matrix estimator}
  \label{fig.distributedResourceAllocation}
\end{figure}

\begin{algorithm}[htb!]
    \caption{PR-CARA algorithm utilizing extended SCI and proactive RSSI estimator in CACC-based platoon driving}
    \begin{algorithmic}[1]
        \STATE Initialize reserved subframe index list $(S^c = \O)$
        \STATE Initialize subsequent transmission resource information list $(S^u = \O)$
        \STATE Initialize CSR list $(CSR_{V^i_{n}} = \O)$
        \STATE Initialize ahead vehicle 1-stage SCI data $(S_{ahead} = \O)$
        \STATE Initialize rear vehicle 1-stage SCI data $(S_{rear} = \O)$
        \STATE Initialize received VUE SCI and TB $(S_{VUE} = \O)$
        \STATE Initialize received PVUE SCI and TB $(S_{PVUE} = \O)$
        \STATE Set the reference signal received power $(RSRP)$ threshold with $-110 dBm$
        \IF{$i ==  0$} \label{lst.line.caccSpecStart}
            \STATE Detect 1-stage SCI data $(S_{rear})$ received from the link $V^{i+1}_n$-to-$V^{i+2}_n$ in the time $\left [t - \tau_{n}, t\right ]$
        \ELSIF{$(2 \leq i \leq (N^p-1))$}
            \STATE Detect 1-stage SCI data $(S_{ahead})$ received from the link $V^{i-1}_n$-to-$V^{i}_n$ in the time $\left [t - \tau_{n}, t\right ]$
            \STATE Detect 1-stage SCI data $(S_{rear})$ received from the link $V^{i+1}_n$-to-$V^{i+2}_n$ in the time $\left [t - \tau_{n}, t\right ]$
        \ELSIF{$i == L$}
            \STATE Detect 1-stage SCI data $(S_{ahead})$ received from the link $V^{i-1}_n$-to-$V^{i}_n$ in the time $\left [t - \tau_{n}, t\right ]$
        \ENDIF \label{lst.line.caccSpecEnd}
        \STATE Detect 1-Stage SCI and TB data $(S_{VUE})$ received from the $n$-th VUE link in the time $\left [t - \tau_{m}, t\right ]$
        \STATE Detect 1-Stage SCI and TB data $(S_{PVUE})$ received from the other PVUE link in the time $\left [t - \tau_{n}, t\right ]$ \label{lst.line.caccSpecPart1}
        \STATE Store data $S_{ahead}$ and $S_{rear}$ in $S^c$ \label{lst.line.caccSpecPart2} 
        \STATE Store data $S_{VUE}$ and $S_{PVUE}$ in $S^u$
        \STATE Store $\left \{c, t \right \}$ resource pair of $(I^l_n(c, t))$ in $CSR_{V^i_{n}}$ $(I^l_n(c, t) \in R^l_n)$
        \IF{$S^c \not= \O$}
            \STATE Remove $\left \{c, t \right \}$ resources pairs transmitting at the reserved subframe index stored in $R_{V^i_{n}}$ from $CSR_{V^i_{n}}$
        \ENDIF
        \STATE Measure RSSI $(\varepsilon^O(c, t))$ of resource at sensing window in the time $[t-\tau_{m}, t-\tau_{m} + \tau_{n}]$
        \IF{$S^u \not= \O$}
            \STATE Set input tuple $(\varepsilon^O(c, t), I^H, D^H, I^E, D^E)$ 
            \STATE Estimate proactive RSSI $\hat{\varepsilon}^P(c, t)$ used estimator $\mathcal{F}(\cdot)$
        \ENDIF
        \WHILE{$\abs{CSR_{V^i_{n}}} \geq 0.2\times \abs{\mathcal{R}^l_n}$} 
            \STATE Store resource pairs $\left \{c, t\right \}$ that satisfy the condition $(\hat{\varepsilon}^P(c, t) \leq RSRP)$ in $CSR_{V^i_{n}}$
            \STATE $RSRP \leftarrow RSRP + 3 dBm$
        \ENDWHILE
        \STATE $V^i_{n}$ randomly selects a resource among the resources with proactive RSSI in the bottom 20\% in $CSR_{V^i_{n}}$ \label{lst.line.csr}
        \STATE Including the subsequent resource information in (RI1, RI2) of the $l$-th packet 1-stage SCI
    \end{algorithmic}\label{alg.proposedMethod}
\end{algorithm}

The proposed algorithm is based on a standardized distributed resource allocation algorithm. Fig. \ref{fig.distributedResourceAllocation} and Algorithm \ref{alg.proposedMethod} describe utilizing the trained estimator in the V2X resource allocation algorithm. Each VUE and PVUE transmit the resource information to be used in the subsequent transmission to the PSCCH of the CAM and PAM, as shown in \raisebox{.9pt}{\textcircled{\raisebox{-.9pt}{1}}} in Fig. \ref{fig.distributedResourceAllocation}. Receiving vehicles must decode the PSCCH's extended SCI. Receiving vehicles can monitor the resources for a subsequent transmission from the adjacent VUE and PVUE based on the decoded SCI information. Each PVUE has a trained estimator that estimates a proactive RSSI based on the monitored subsequent transmission information as shown in \raisebox{.9pt}{\textcircled{\raisebox{-.9pt}{2}}}. Five features of the data, $\left \{ \varepsilon^O_{n,l}, I^{H}, D^{H}, I^{E}, D^{E} \right \}$, were input into the estimator. The output of the estimator is the proactive RSSI $(\hat{\varepsilon}^P_{n,l})$, which reflects subsequent transmission information. Here, $\varepsilon$ represents the actual (measured) RSSI and $\hat{\varepsilon}$ represents the estimated RSSI. Subsequently, the proactive resource allocation algorithm \raisebox{.9pt}{\textcircled{\raisebox{-.9pt}{3}}} is executed based on the proactive RSSI.

This proactive resource allocation is performed using Algorithm \ref{alg.proposedMethod}. This represents the proactive resource allocation process performed when the $n$-th PVUE $(V^i_{n})$ transmits the $l$-th packet. Here, $(V^i_{n})$ is the $i$-th vehicle in the platoon $(i=0, 1, 2, ..., N^p)$, and $V^0_n$ is the platoon leader vehicle, $N^p$ is the total number of vehicles per platoon. PVUE $(V^i_{n})$ selects one resource from $CSR_{V^i_{n}}$ formed through this process in line \ref{lst.line.csr} of Algorithm \ref{alg.proposedMethod} and transmits the $l$-th packet. In a CACC-based platoon driving service, the primary communication targets are the lead and rear vehicles. Accordingly, lines \ref{lst.line.caccSpecStart}–\ref{lst.line.caccSpecEnd} of Algorithm \ref{alg.proposedMethod} store the SCI data ($S_{ahead}$ and $S_{rear}$) from the lead and rear vehicles in dataset $S^c$ to prevent packet collisions. In other words, $S^c$ must include the 1-stage SCI information of the primary communication targets. Therefore, Lines \ref{lst.line.caccSpecStart}–\ref{lst.line.caccSpecEnd} can be adjusted based on the applied service.

The resource selection scheme considers two methods to explore the optimal availability of the estimated RSSI. The first method is to assess the proactive sensing matrix and selects a resource with the minimum RSSI, similar to the optimal solution in (\ref{eq.transformedOptimizationProblem}). The second method is to include resources with an estimated proactive RSSI value in the lower 20\% of CSR, such as in Algorithm \ref{alg.proposedMethod}, considering the possibility that the SCI of an adjacent vehicle will not be decoded. The two methodologies applied to the proposed algorithm are compared in terms of communication QoS in Section \ref{sec.simulationResult}. The results of these two methodologies provide insight for developing resource allocation methods that monitor subsequent resource information.

\section{Simulation Results}\label{sec.simulationResult}
The proposed proactive resource allocation algorithm can be applied without distinguishing between periodic and aperiodic services. In this section, the conventional SB-SPS and SB-DS methods are sequentially compared with the proposed algorithm. In the periodic message scenario, the communication reliability and IPG QoS were used as performance indicators. In the aperiodic message scenario, the processing times for cooperative services and event-triggered messages were set as performance indicators for analysis.

\begin{figure}[t!]
  \centering
  \includegraphics[clip, width = 1\columnwidth]{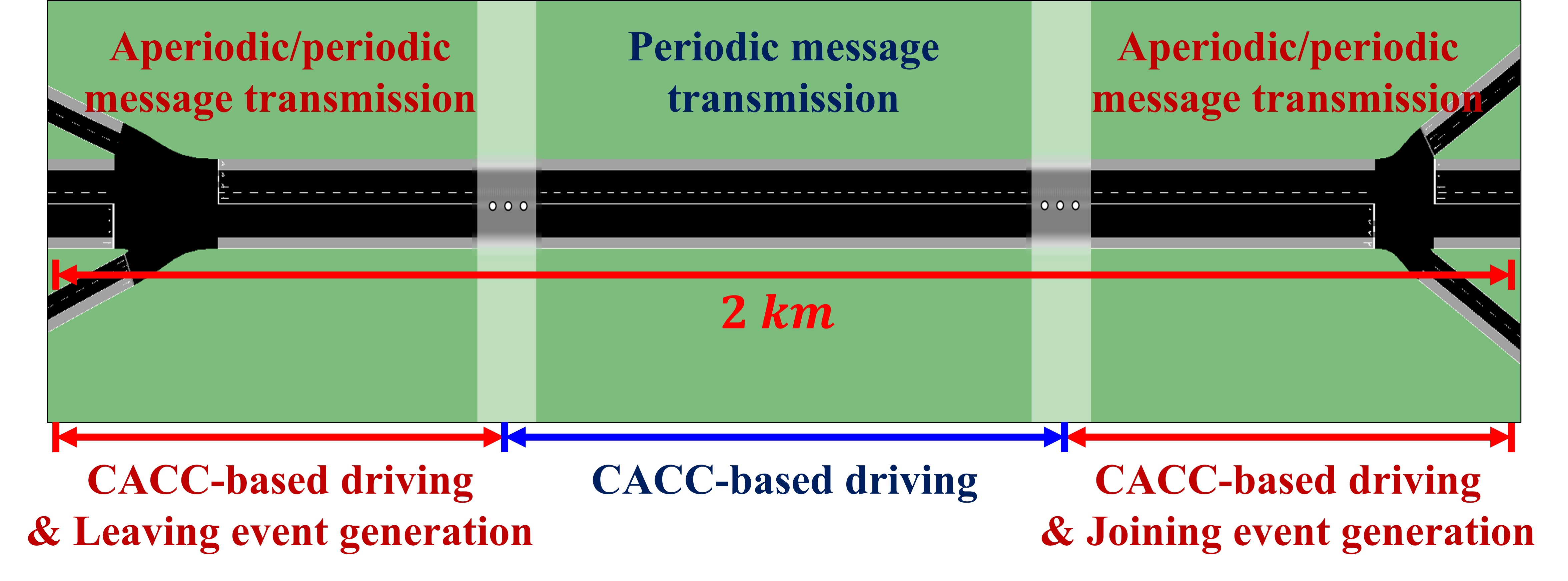}
  \caption{Road configuration implemented in SUMO for algorithm verification}
  \label{fig.sumoRoadScenario}
\end{figure}

\subsection{Simulation and Scenario Setup}
The proposed resource allocation algorithm is verified by integrating the SUMO traffic simulator \cite{SUMO} and modified WiLabV2Xsim\cite{wilabV2Xsim}. The SUMO traffic simulator is a free open-source traffic simulator that supports the modeling of complex transportation systems, including road vehicles, public transportation, and pedestrians. The highway traffic flow generated by the SUMO traffic simulator reflects vehicle congestion and terrain, enabling a realistic traffic flow implementation. This enables the verification of the wireless resource allocation algorithm in a more realistic road traffic environment compared with using only WiLabV2Xsim.

The road environment for the algorithm verification is shown in Fig. \ref{fig.sumoRoadScenario}. The center section is where CACC-based platoon driving is performed, and resources for periodic messages (PAMs) are allocated using the proposed algorithm. In the left section, leaving events occur when vehicles leave the platoon, whereas in the right section, joining events occur when vehicles join the platoon. The messages generated in these sections include not only periodic PAM for CACC services but also event-triggered messages with aperiodic characteristics. Both CACC-based platoon driving and joining/leaving event traffic are implemented in the SUMO simulator found in the GitHub repository\footnote{\url{https://github.com/Hanyang-CAM-Lab/SUMO_Matlab_Cosimulator_event_Triggered_Scenario}}.

Simulation and wireless communication parameters for verification are shown in Table \ref{tbl.simulParams}. It should be noted in the organized table that PAM and event-triggered messages require shorter transmission time intervals and narrower resource selection windows. This causes high packet collisions and decoding errors, and the proposed algorithm is verified under these challenging conditions.

To demonstrate the superiority of the proposed algorithm, it is compared with the following algorithms: First, the standardized SB-SPS method by \cite{sbsps}, which maintains the same resource for a configured SPS period; second, SB-DS, which, unlike SB-SPS, reallocates resources for each transmission, partially mitigating SINR degradation caused by SPS characteristics; 
Third, based on the core idea of Ali et al. \cite{extendedDataAli1, extendedDataAli2, extendedDataAli3} and Sabeeh et al. \cite{extendedDataSabeeh1, extendedDataSabeeh2, extendedDataSabeeh3}, it is an algorithm that shares subframe offset information and utilizes it to avoid packet collisions. These algorithms adopt a similar approach to our study and serve as the primary comparison targets; fourth, an algorithm that selects the optimal solution to the minimum RSSI detection problem, a transformed optimization problem (\ref{eq.transformedOptimizationProblem}), using proactive RSSI; and fifth, the PR-CARA algorithm, which randomly selects one of the resources from the CSR constructed from the lowest 20\% RSSI values, as shown in Algorithm \ref{alg.proposedMethod}. Each of these algorithms is executed and compared in the mentioned road environment.

\begin{table}[]
\caption{Simulation parameters}
\label{tab:my-table}
\resizebox{\columnwidth}{!}{%
\begin{tabular}{|c|ccc|}
\hline
Name                                                                                   & \multicolumn{1}{c|}{VUE}                                                                  & \multicolumn{2}{c|}{PVUE}                                                                                                                                                 \\ \hline
Packet name                                                                            & \multicolumn{1}{c|}{CAM}                                                                  & \multicolumn{1}{c|}{PAM}                                                                 & Event packet                                                         \\ \hline
Payload size                                                                           & \multicolumn{1}{c|}{300 Byte}                                                             & \multicolumn{1}{c|}{500 Byte}                                                            & 500 Byte                                                                       \\ \hline
Transmission rate                                                                      & \multicolumn{1}{c|}{10 Hz}                                                                & \multicolumn{1}{c|}{50 Hz}                                                               & -                                                                              \\ \hline
TTI                                                                                    & \multicolumn{1}{c|}{100 ms}                                                               & \multicolumn{1}{c|}{20 ms}                                                               & -                                                                              \\ \hline
MCS                                                                                    & \multicolumn{1}{c|}{3}                                                                    & \multicolumn{1}{c|}{3}                                                                   & 3                                                                              \\ \hline
Selection window                                                                          &
\multicolumn{1}{c|}{[1, 100] ms}                                                           & 
\multicolumn{1}{c|}{[1, 20] ms}                                            & [1, 20] ms
                                                                        \\ \hline
Vehicle density $(\rho)$                                                               & \multicolumn{3}{c|}{$\rho$ = [40, 80, 120, ..., 400]}                                                                                                                                                                                                                 \\ \hline
\# of total vehicle                                                                     & \multicolumn{3}{c|}{[80, 160, 240, ..., 800]}                                                                                                                                                                                                                         \\ \hline
Channel model                                                                          & \multicolumn{3}{c|}{Winner+B1}                                                                                                                                                                                                                                        \\ \hline
Bandwidth                                                                              & \multicolumn{3}{c|}{20 MHz}                                                                                                                                                                                                                                           \\ \hline
Antenna gain $(G_{tx}, G_{rx})$                                                        & \multicolumn{3}{c|}{3 dBi}                                                                                                                                                                                                                                            \\ \hline
Noise figure                                                                           & \multicolumn{3}{c|}{9 dB}                                                                                                                                                                                                                                             \\ \hline
Simulation time per 1-sample                                                           & \multicolumn{3}{c|}{30 s}                                                                                                                                                                                                                                             \\ \hline
\# of Monte Carlo simulation sample                                                     & \multicolumn{3}{c|}{\begin{tabular}[c]{@{}c@{}}100 [\# of simulation sample per unit $\rho$]\\ 10 [\# of $\rho$ case]\\ 1500 [\# of transmission packet per 1-sample]\\ 5 [\# of PVUE]\\ \# of total sample: 100 $\times$ 10 $\times$ 1500 $\times$ 5\end{tabular}} \\ \hline
\end{tabular}
}\label{tbl.simulParams}
\end{table}

\begin{figure*}[htb!]
  \centering
  \includegraphics[clip, width = 1\textwidth]{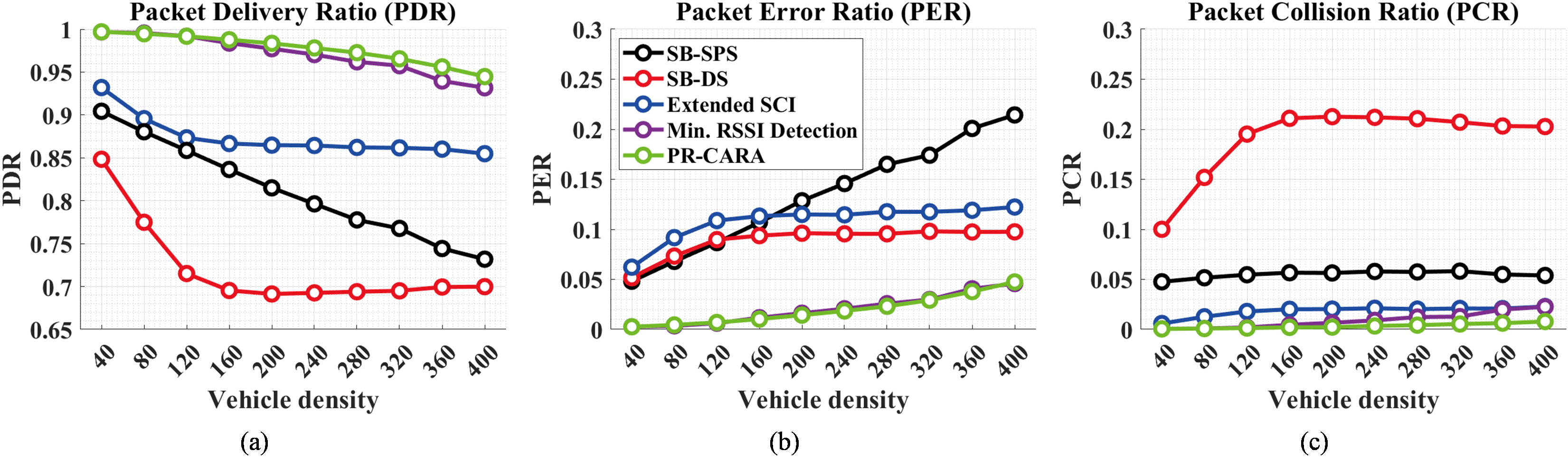}
  \caption{(a) Packet delivery ratio; (b) packet error ratio; (c) packet collision ratio of transmitted PAM packet in CACC-based platoon driving scenario.}
  \label{fig.caccReliability}
\end{figure*}

\subsection{Verification in Periodic Message Scenario}
This section addresses the QoS of the PAM communication links transmitted by vehicles operating a CACC-based platoon driving service. The main performance indicators were $PDR$, packet error ratio ($PER$), packet collision ratio ($PCR$), and $IPG$. $PDR$, $PER$, and $PCR$ are related to the communication reliability, whereas $IPG$ is related to the latency. $PDR$, $PER$, and $PCR$ were calculated using the following formula:
\begin{subequations}
        \begin{equation}
        \begin{aligned}
            PDR & = & \frac{\sum_{n\in\mathcal{N}} \sum_{l\in\mathcal{L}}\mathds{1}^{Reception}_{\left \{\gamma^l_n \geq  \gamma_0\right \}}(c_n^l, t_n^l)}{\sum_{n\in\mathcal{N}} \sum_{l\in\mathcal{L}}I_n^l(c_n^l, t_n^l)},\\
        \end{aligned}
        \end{equation}
        \begin{equation}
        \begin{aligned}
            PER & = & \frac{\sum_{n\in\mathcal{N}} \sum_{l\in\mathcal{L}}\mathds{1}^{Error}_{\left \{\gamma^l_n <  \gamma_0\right \}}(c_n^l, t_n^l)}{\sum_{n\in\mathcal{N}} \sum_{l\in\mathcal{L}}I_n^l(c_n^l, t_n^l)},\\
        \end{aligned}
        \end{equation}
        \begin{equation}
        \begin{aligned}
            PCR & = & \frac{\sum_{n\in\mathcal{N}} \sum_{l\in\mathcal{L}}\mathds{1}^{Collision}_{\left \{ t_n^l = t_{n.target}^l \right \}}(c_n^l, t_n^l)}{\sum_{n\in\mathcal{N}} \sum_{l\in\mathcal{L}}I_n^l(c_n^l, t_n^l)},\\
        \end{aligned}
        \end{equation}\label{eq.reli}
\end{subequations}
 where $(c_n^l, t_n^l)$ represent the subchannel and subframe indices used by the $n$-th PVUE in the $l$-th transmission. In addition, the sum of $PDR$, $PER$, and $PCR$ is $1$, and the Reception, Error, and Collision in $\mathds{1}$ refer to the packet transmission status.  This is defined as $\mathds{1}^{Reception}$ when a packet is received successfully, $\mathds{1}^{Error}$ when a packet is not decoded because of the deterioration of the SINR, and $\mathds{1}^{Collision}$ when a packet is transmitted in the same subframe, resulting in an HD error. Accordingly, $t^l_{n.target}$ is the subframe index used by the unicast communication target of the $n$-th PVUE. The indicator functions are expressed as follows:
\begin{subequations}
    \begin{equation}
        \mathds{1}^{Reception}(c_n^l, t_n^l)=\begin{cases}
    \begin{aligned}
         1 \quad & \text{if} \quad \gamma_{n}^l(c_n^l, t_n^l) \geq \gamma^l_{n, 0}\\
         0 \quad & otherwise.
    \end{aligned}
    \end{cases}
    \end{equation}
    \begin{equation}
        \mathds{1}^{Error}(c_n^l, t_n^l)=\begin{cases}
    \begin{aligned}
         1 \quad & \text{if} \quad \gamma_{n}^l(c_n^l, t_n^l) < \gamma^l_{n, 0}\\
         0 \quad & otherwise.
    \end{aligned}
    \end{cases}
    \end{equation}
    \begin{equation}
        \mathds{1}^{Collision}(c_n^l, t_n^l)=\begin{cases}
    \begin{aligned}
         1 \quad & \text{if} \quad t_n^l = t_{n.target}^l\\
         0 \quad & otherwise.
    \end{aligned}
    \end{cases}
    \end{equation}\label{eq.reliIndicator}
\end{subequations}
Finally, $PDR$, $PER$, and $PCR$ in the CACC-based platoon driving environment, as calculated using the above equations, are shown in Fig. \ref{fig.caccReliability}.

The black solid line represents the results of the SB-SPS algorithm, whereas the red solid line represents those of the SB-DS algorithm. In terms of $PDR$, it is evident that SB-SPS performs better than SB-DS. However, a more meaningful interpretation is possible by separating the overall packet status into the error status owing to SINR deterioration and the collision status owing to HD characteristics. First, in terms of error status, SB-SPS shows a higher $PER$ at higher congestion than SB-DS. However, in terms of the collision status, SB-DS showed a much higher $PCR$. These results reflect both the advantages of SB-DS in securing an excellent transport packet SINR and the disadvantages of increasing the probability of packet collisions.

The blue solid line represents the result of the algorithm that uses the extended 1-Stage SCI solely for packet collision avoidance. This is the core approach of \cite{extendedDataAli1, extendedDataAli2, extendedDataAli3, extendedDataSabeeh1, extendedDataSabeeh2, extendedDataSabeeh3, tsnamVTC} and aims to minimize the disadvantages of SB-DS. This algorithm focuses exclusively on avoiding packet collisions by monitoring the resources that the key communication targets plan to use in subsequent transmissions. Consequently, it is confirmed that $PCR$ can be minimized by addressing the packet collision problem of SB-DS, thereby improving $PDR$.

The purple and green solid lines represent the results of applying the proactive resource allocation algorithm using the extended SCI and proactive RSSI estimators. This approach not only avoids packet collisions but also enables optimal resource selection, which ensures a high SINR for the transmitted packets. As a result, compared to the other algorithms, it exhibits lower $PER$ and $PCR$ while demonstrating a significantly improved $PDR$. Consequently, the proposed system and algorithm greatly enhance the communication reliability.

The purple solid line represents the result of the quasi-optimal solution of the minimum RSSI detection problem defined in (\ref{eq.transformedOptimizationProblem}). The green solid lines represent the results obtained using the PR-CARA algorithm. The results show that the PR-CARA algorithm (green solid line) offers better communication reliability than the quasi-optimal minimum RSSI resource selection method (purple solid line). This improvement is attributed to the decoding uncertainty of the extended SCI. The proposed algorithm estimates the proactive RSSI value when the extended SCI of adjacent vehicles is decoded. However, decoding the extended SCI cannot be guaranteed under high traffic congestion, which hinders the resource monitoring process between adjacent vehicles. Without prior resource monitoring, the high similarity in RSSI values for the same resource increases the likelihood of packet collisions. This is confirmed by the $PCR$ values for the green and purple lines, which increase steadily as the vehicle traffic congestion $\rho$ approaches 400. These results suggest that the PR-CARA algorithm (green line), which considers the decreased probability of decoding extended SCI owing to increased traffic congestion, represents an optimal strategy for utilizing proactive RSSI.

\begin{figure}[tb!]
\centering
\subfloat[]{
\includegraphics[clip, width = 0.8\columnwidth]{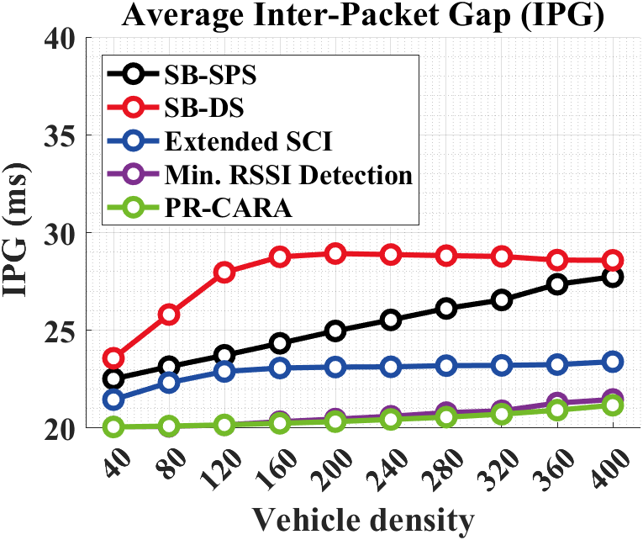}
}

\subfloat[]{
\includegraphics[clip, width = 0.8\columnwidth]{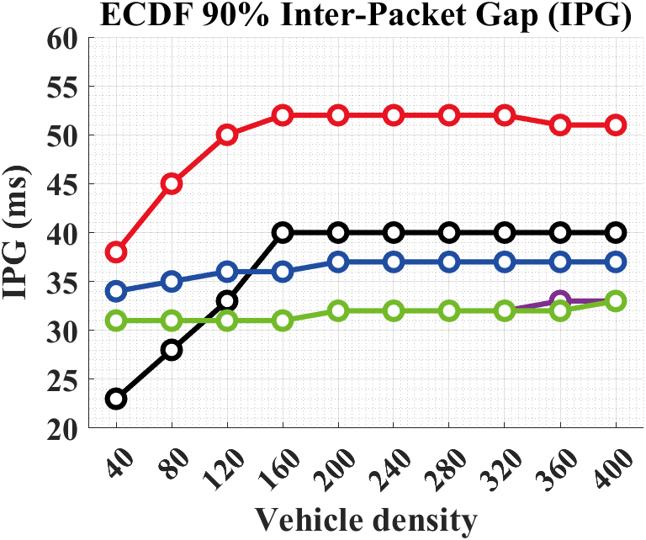}
}
\caption{(a) Average IPG of transmitted PAM packet in CACC-based platoon driving scenario; (b) 90th percentile value of IPG ECDF in CACC-based platoon driving scenario}
\label{fig.ipg}
\end{figure}

Next, we present the IPG results. The IPG, which is closely related to the latency, is calculated as follows:
\begin{subequations}
    \begin{equation}
        g^* = \underset{g}{argmin}\left| \mathds{1}^{Reception}(c_n^l, t_n^l) - \mathds{1}^{Reception}(c_n^{l+g}, t_n^{l+g}) \right|,
    \end{equation}
    \begin{equation}
        IPG = \frac{1}{N\cdot L} \sum_{n\in\mathcal{N}}\sum_{l\in\mathcal{L}} \left|t_n^{l+g^*} - t_n^{l}\right|.
    \end{equation}
\end{subequations}
$IPG$ is the time interval between packets that were successfully received. This is an important performance indicator that describes the latency and determines $\theta_c$ in Fig. \ref{fig.dynamicsParamOfCacc}. The variation in the average $IPG$ and 90th percentile values of $IPG$ empirical cumulative distribution function (ECDF) with vehicle density is shown in Fig. \ref{fig.ipg}.

\begin{figure}[htb!]
\centering
\subfloat[]{
\includegraphics[clip, width = 0.8\columnwidth]{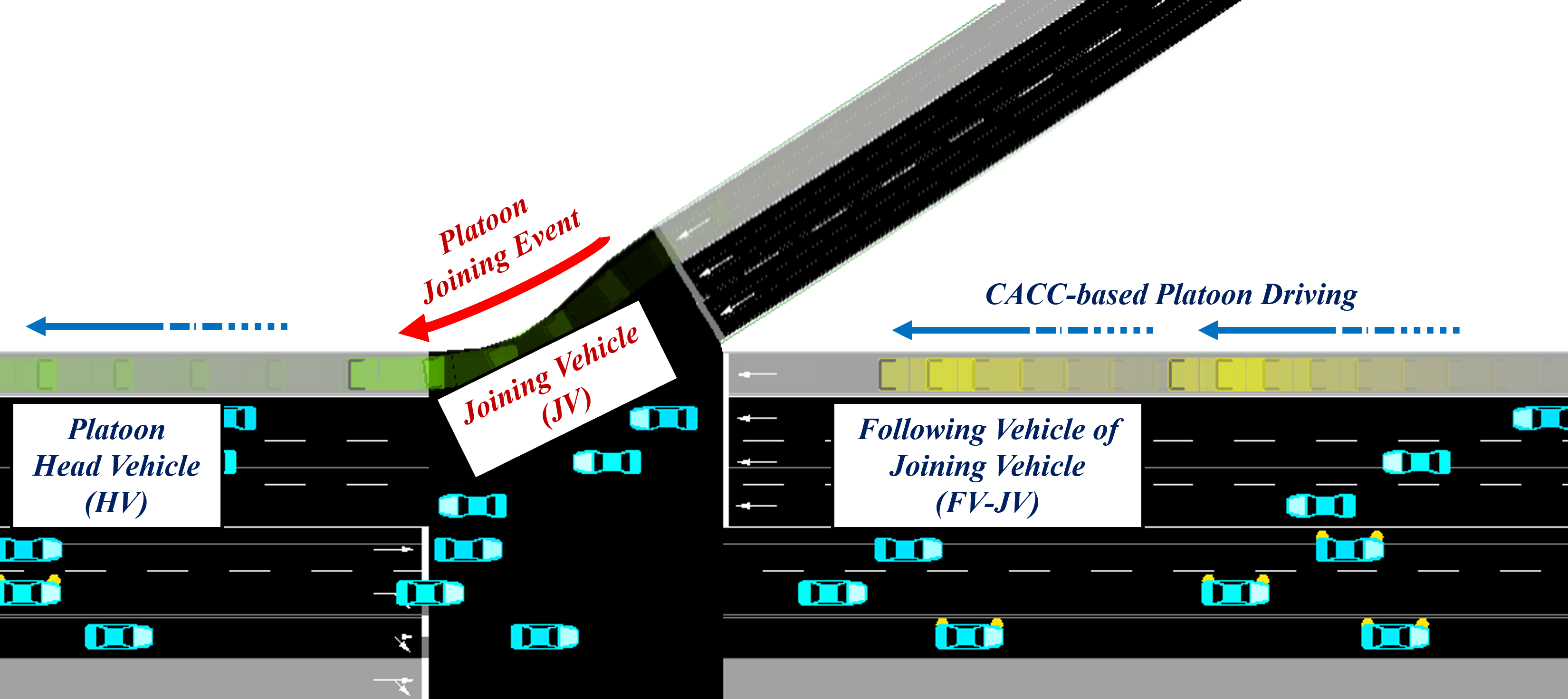}
}

\subfloat[]{
\includegraphics[clip, width = 0.8\columnwidth]{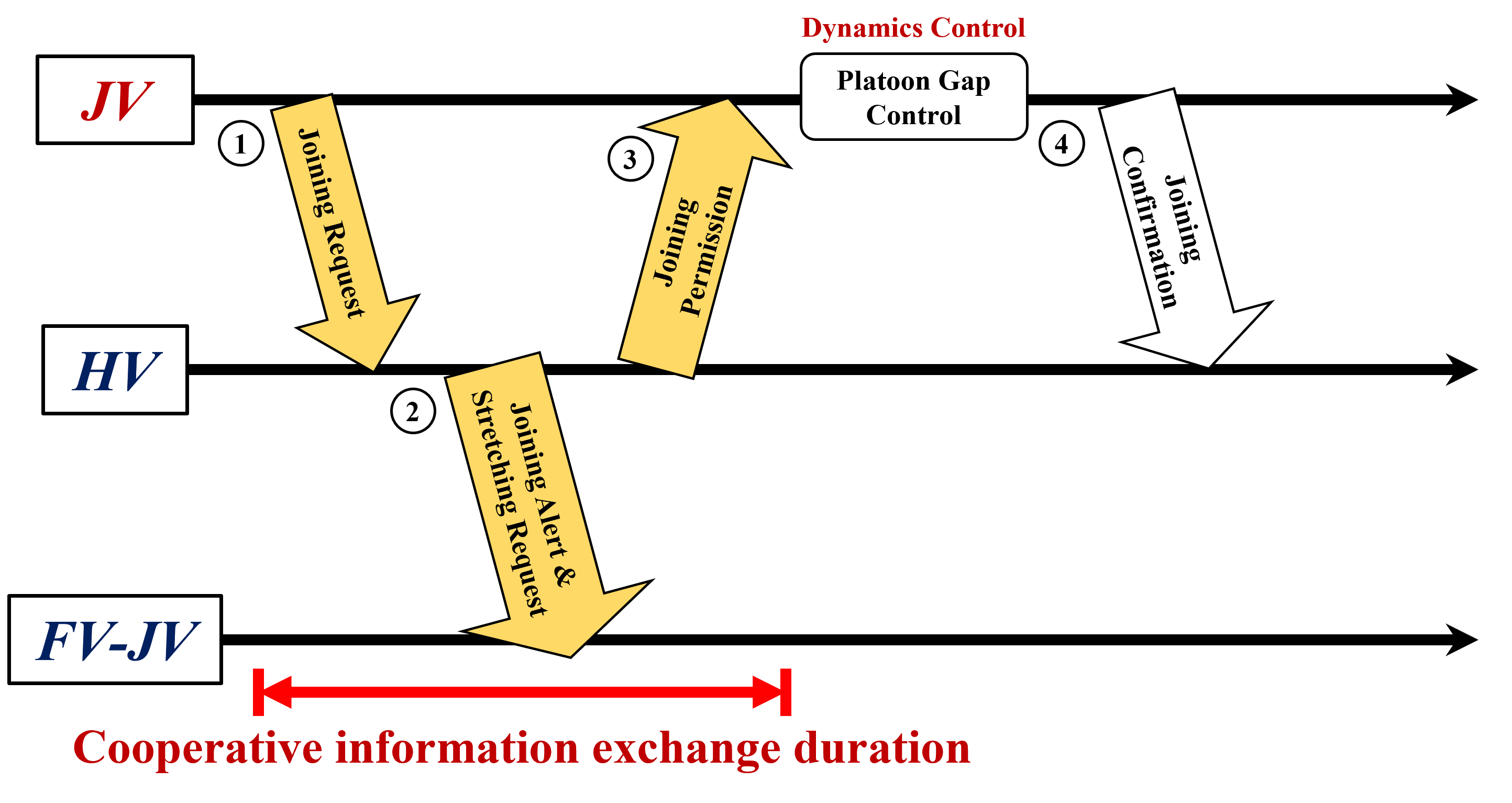}
}
\caption{(a) Joining event-triggered traffic configuration; (b) information exchange and dynamics control protocol}
\label{fig.joiningProcedure}
\end{figure}
The $IPG$ performance exhibits a similar trend to reliability. As shown in Fig. \ref{fig.ipg}-(a), the extended SCI algorithm effectively mitigates packet collisions, leading to improved $PDR$ and, consequently, a notable enhancement in $IPG$. The proactive RSSI-based resource allocation method (purple and green solid lines) also demonstrates that $IPG$ is close to 20 ms, with the PR-CARA algorithm (green solid line) showing the best performance. However, the 90th percentile of the IPG ECDF in Fig. \ref{fig.ipg}-(b) presents a different pattern. This is the value where 90\% of the IPG values are satisfied. Unlike the previous results, SB-SPS demonstrates superior performance in low-congestion traffic. At the same time, the proposed proactive RSSI-based resource allocation method (purple and green solid lines) outperforms in high-traffic congestion conditions. This difference arises from the characteristics of the SPS method. SPS reuses the set resources for a certain period and transmits at every $RRI$ until the RC becomes 0. As a result, low IPG values are observed in low-congestion traffic conditions where PDR is guaranteed. However, in high-congestion traffic, the PDR of SB-SPS cannot be guaranteed, resulting in higher IPG values compared to the proposed algorithm. Consequently, in low-congestion traffic, SB-SPS can maintain a small $\theta_c$, but in high-congestion traffic, where communication reliability cannot be guaranteed, the proposed proactive RSSI-based resource allocation algorithm maintains a small $\theta_c$, ensuring a stable CACC-based platoon driving service.

\subsection{Verification in Aperiodic Message Scenario}
The algorithm is verified for the aperiodic message scenario during a joining/leaving event. The information exchange and control procedures during the joining/leaving event are explained \cite{joiningLeaving}. Such cooperative services require short processing times, and the processing times for cooperative services and event-triggered messages are presented as performance indicators.

\subsubsection{Joining event-triggered message scenario}
The operation and service flow of the platooning vehicles during the joining event, according to the information exchange and control procedures, are shown in Fig. \ref{fig.joiningProcedure}. Fig. \ref{fig.joiningProcedure}-(a) shows a simulation of a joining event, where information exchange and dynamics control protocols are mainly performed in the platoon head vehicle (HV), leaving vehicle (LV), and the following vehicle of the leaving vehicle (FV-LV). Fig. \ref{fig.joiningProcedure}-(b) depicts the protocol for information exchange and dynamics control. This procedure is divided into four stages: joining request, joining alert, stretching request, joining permission, and joining confirmation, as described by \cite{joiningLeaving}.

\begin{figure}[tb!]
\centering
\subfloat[]{
\includegraphics[clip, width = 0.8\columnwidth]{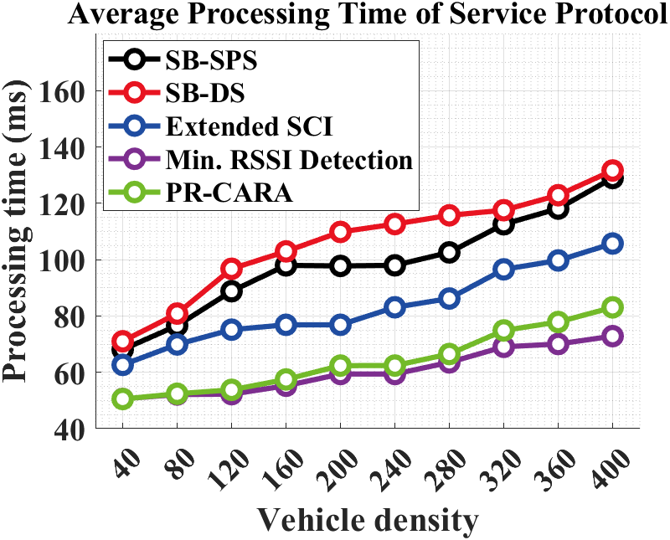}
}

\subfloat[]{
\includegraphics[clip, width = 0.8\columnwidth]{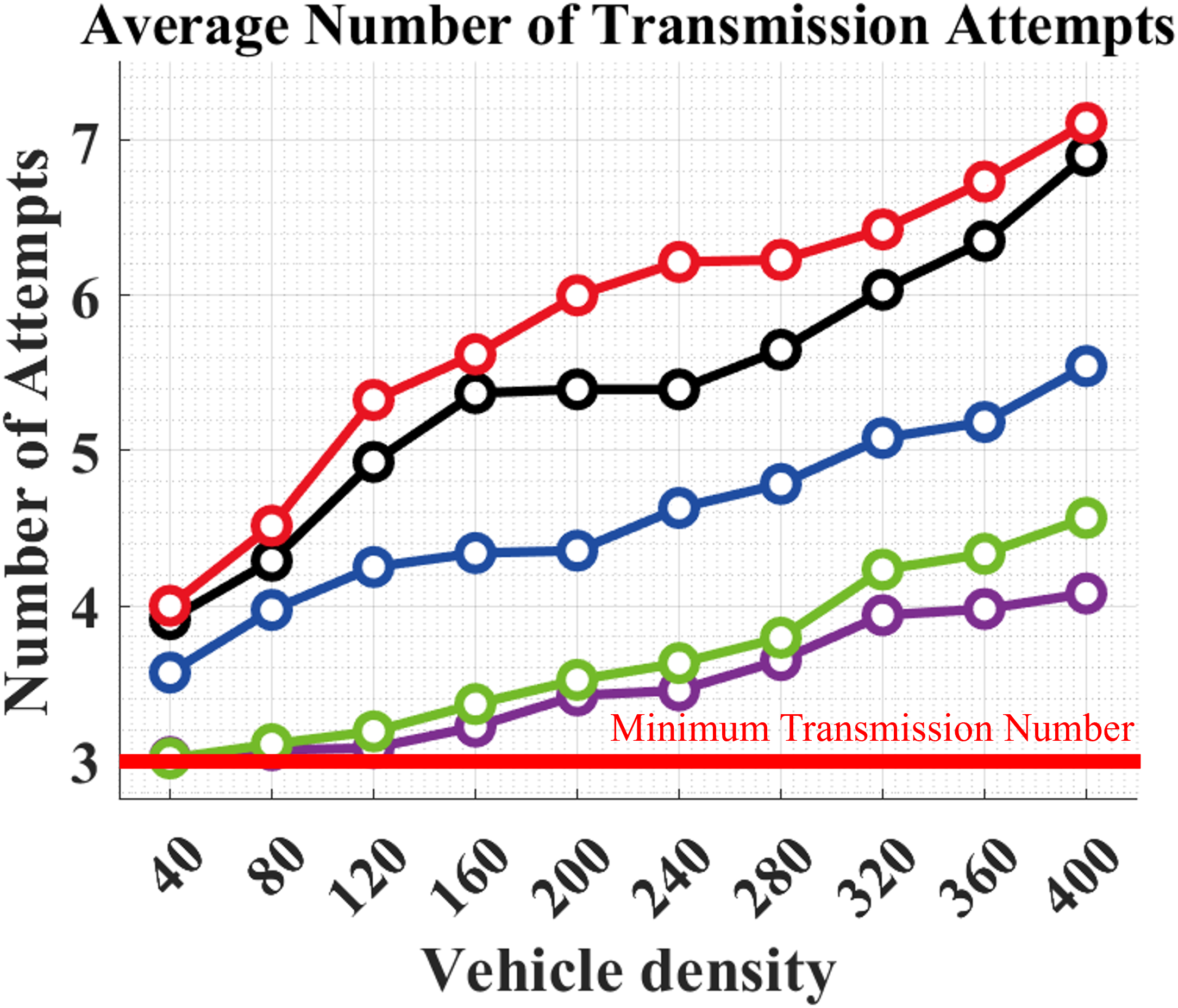}
}
\caption{Joining event protocol (a) average processing time and (b) average number of transmission attempts for vehicle density variation}
\label{fig.joiningEventProtocolResult}
\end{figure}

This section uses the processing time of V2V communication protocols before dynamic control as a performance indicator. This is referred to as the cooperative information exchange duration. The proposed algorithm aims to reduce the protocol processing time. The performance of the proposed method is illustrated in Fig. \ref{fig.joiningEventProtocolResult}. Fig. \ref{fig.joiningEventProtocolResult}-(a) illustrates the processing time during cooperative information exchange in the joining event, where shorter times indicate better performance. Fig. \ref{fig.joiningEventProtocolResult}-(b) depicts the number of transmission attempts until the end of the event protocol, where a value closer to 3, the minimum number of transmissions, indicates better performance. The minimum RSSI detection algorithm performed best for joining events (event-triggered aperiodic messages).

Remarkably, unlike the CACC-based platoon driving (periodic message) scenario, the relative effectiveness of the minimum RSSI detection (purple solid line) and PR-CARA algorithm (green solid line) is reversed. This result can be explained by the locations of the vehicles performing the joining event. A joining event is performed in the merging section before the JV enters the highway, as shown in Fig. \ref{fig.joiningProcedure}-(a).Thus, the channel congestion experienced by HV and FV-JV located on the highway increases linearly with vehicle density; however, the channel congestion experienced by JV is maintained. This leads to better performance for the minimum RSSI detection algorithm that performs greedy resource allocation compared to the PR-CARA algorithm.

\subsubsection{Leaving event-triggered message scenario}
\begin{figure}[tb!]
\centering
\subfloat[]{
\includegraphics[clip, width = 0.8\columnwidth]{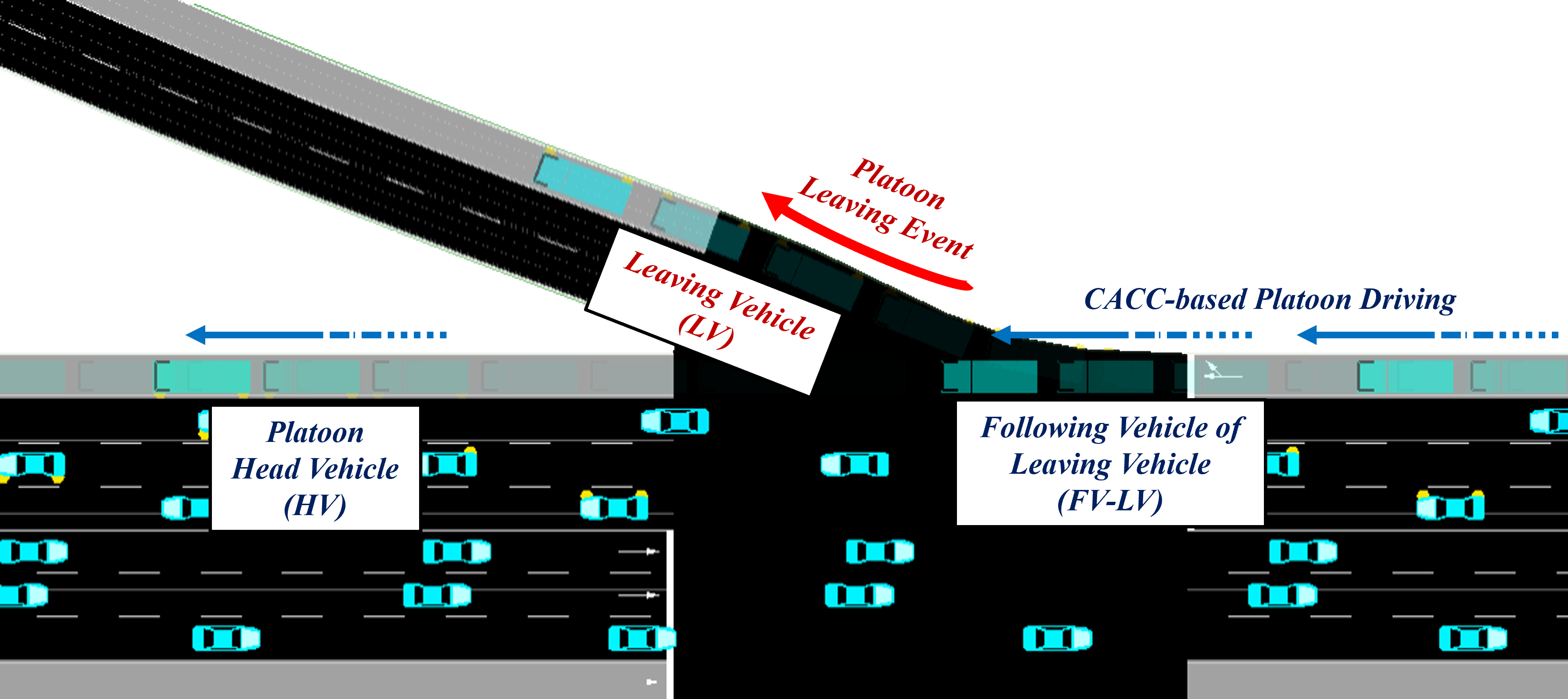}
}

\subfloat[]{
\includegraphics[clip, width = 0.8\columnwidth]{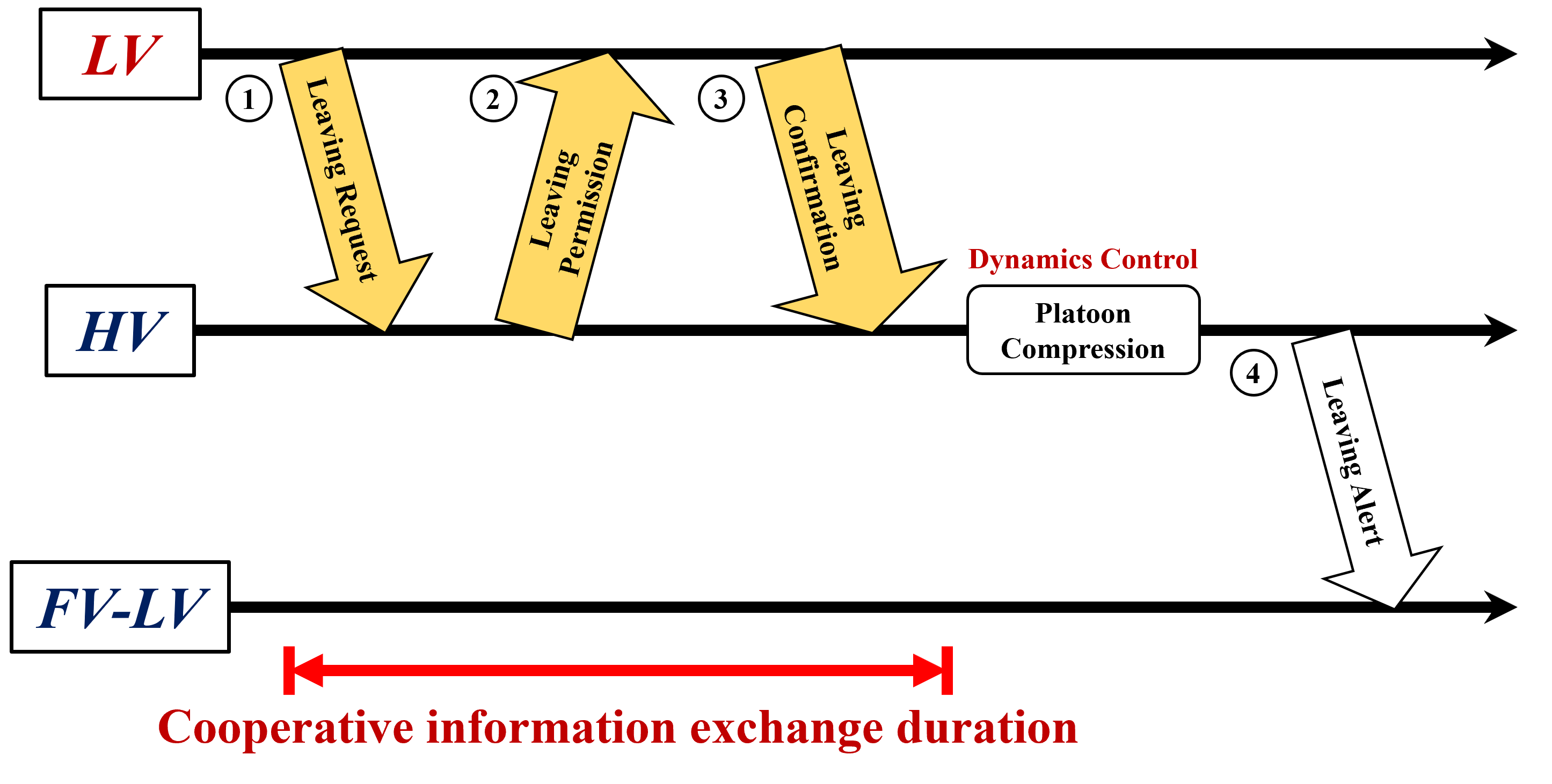}
}
\caption{(a) Leaving event triggered traffic configuration; (b) information exchange and dynamics control protocol}
\label{fig.leavingProcedure}
\end{figure}

Similarly, when the leaving event is performed, the operation and service flow according to the information exchange  and control procedures of the platooning vehicle can be expressed as shown in Fig. \ref{fig.leavingProcedure}. Fig. \ref{fig.leavingProcedure}-(a) shows a simulation of a leaving event, and information exchange and dynamics control protocols are mainly performed by the illustrated platoon head vehicle (HV), leaving vehicle (LV), and following vehicle of the leaving vehicle (FV-LV). Fig. \ref{fig.leavingProcedure}-(b) depicts the information exchange and dynamics control protocol. The procedure is divided into four stages: leaving requests, permission, confirmation, and alertness.

\begin{figure}[htb!]
\centering
\subfloat[]{
\includegraphics[clip, width = 0.8\columnwidth]{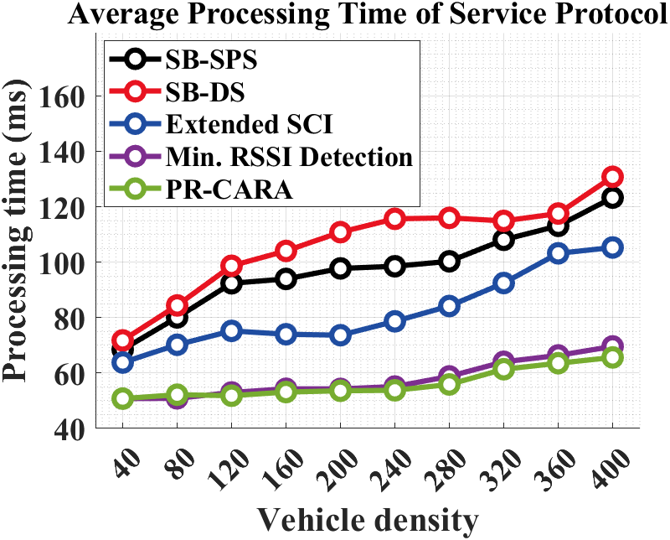}
}

\subfloat[]{
\includegraphics[clip, width = 0.8\columnwidth]{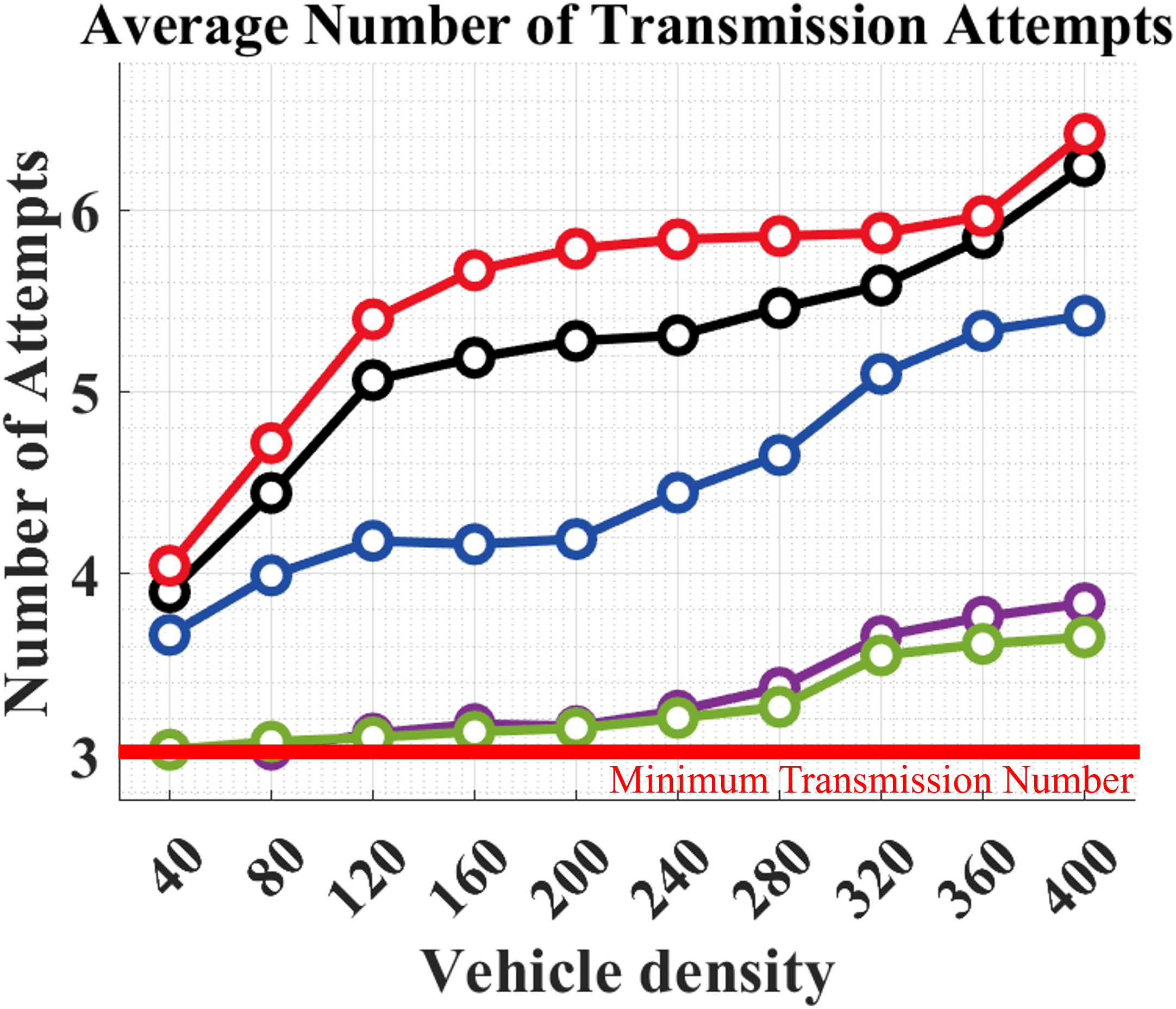}
}
\caption{Leaving event protocol (a) average processing time and (b) average number of transmission attempts for vehicle density variation}
\label{fig.leavingEventProtocolResult}
\end{figure}

Fig. \ref{fig.leavingEventProtocolResult} shows (a) the average process time required and (b) the average number of transmission attempts during the cooperative information exchange in a leaving event. The PR-CARA algorithm exhibited the best results for leaving events (event-triggered aperiodic messages). This result is similar to that of the CACC-based platoon driving (periodic message) scenario. This trend can be attributed to the location of the vehicles performing the leaving event. The leaving event is performed on a highway before entering a branch section of the LV, as shown in Fig. \ref{fig.leavingProcedure}-(a). HV, LV, and FV-LV are all on the highway, so they experience the same level of channel congestion. Consequently, it shows the same performance pattern as that of the CACC-based platoon driving(periodic message service) analysis.

\section{Conclusion}\label{sec.conclusion}
This paper presented a resource allocation algorithm for V2X-based fully autonomous driving that focuses on improving the communication QoS of URLLC services. Leveraging the extended 1-stage SCI and deep learning-based proactive RSSI estimator, the proposed algorithm, PR-CARA, solves hidden/exposed node problems that cause packet collisions and transmission errors and improves communication reliability, latency, and processing time compared to conventional methods. Simulation results reveal that PR-CARA performs well for both periodic and aperiodic message services. These findings highlight the versatility of PR-CARA, as it is not limited to specific services but can be flexibly applied to various V2X applications. Moreover, by addressing fundamental issues in distributed resource allocation, such as RSSI inaccuracy, using physics-based AI, PR-CARA provides a robust and practical solution that enhances communication QoS across various scenarios, paving the way for more reliable and scalable autonomous driving services. However, this study focused on improving the detection process, which suggests further potential for performance enhancement through improvement of the selection process. Accordingly, future studies should aim to develop solutions to improve the selection process. Specifically, the goal could be to integrate advanced RL-based method with physics-based AI to develop algorithms capable of providing practical solutions across various environments.

\bibliographystyle{IEEEtran}
\bibliography{TITS_Paper_v8_arXiv}

\begin{thebibliography}{10}
\providecommand{\url}[1]{#1}
\csname url@samestyle\endcsname
\providecommand{\newblock}{\relax}
\providecommand{\bibinfo}[2]{#2}
\providecommand{\BIBentrySTDinterwordspacing}{\spaceskip=0pt\relax}
\providecommand{\BIBentryALTinterwordstretchfactor}{4}
\providecommand{\BIBentryALTinterwordspacing}{\spaceskip=\fontdimen2\font plus
\BIBentryALTinterwordstretchfactor\fontdimen3\font minus \fontdimen4\font\relax}
\providecommand{\BIBforeignlanguage}[2]{{%
\expandafter\ifx\csname l@#1\endcsname\relax
\typeout{** WARNING: IEEEtran.bst: No hyphenation pattern has been}%
\typeout{** loaded for the language `#1'. Using the pattern for}%
\typeout{** the default language instead.}%
\else
\language=\csname l@#1\endcsname
\fi
#2}}
\providecommand{\BIBdecl}{\relax}
\BIBdecl

\bibitem{citsTITS1}
M.~Brambilla, M.~Nicoli, G.~Soatti, and F.~Deflorio, ``Augmenting vehicle localization by cooperative sensing of the driving environment: Insight on data association in urban traffic scenarios,'' \emph{IEEE Transactions on Intelligent Transportation Systems}, vol.~21, no.~4, pp. 1646--1663, 2020.

\bibitem{citsTITS2}
Z.~Huang, S.~Chen, Y.~Pian, Z.~Sheng, S.~Ahn, and D.~A. Noyce, ``Toward c-v2x enabled connected transportation system: Rsu-based cooperative localization framework for autonomous vehicles,'' \emph{IEEE Transactions on Intelligent Transportation Systems}, vol.~25, no.~10, pp. 13\,417--13\,431, 2024.

\bibitem{citsTITS3}
A.~H. Sodhro, S.~Pirbhulal, G.~H. Sodhro, M.~Muzammal, L.~Zongwei, A.~Gurtov, A.~R.~L. de~Macêdo, L.~Wang, N.~M. Garcia, and V.~H.~C. de~Albuquerque, ``Towards 5g-enabled self adaptive green and reliable communication in intelligent transportation system,'' \emph{IEEE Transactions on Intelligent Transportation Systems}, vol.~22, no.~8, pp. 5223--5231, 2021.

\bibitem{citsTITS4}
F.~Abbas, P.~Fan, and Z.~Khan, ``A novel low-latency v2v resource allocation scheme based on cellular v2x communications,'' \emph{IEEE Transactions on Intelligent Transportation Systems}, vol.~20, no.~6, pp. 2185--2197, 2019.

\bibitem{cv2x}
``Evolved universal terrestrial radio access(e-utra) and evolved universal terrestrial radio access network (e-utran): Overall description: Stage 2,'' \emph{3GPP TR 36.300}, 2017.

\bibitem{nrv2x}
``Study on nr vehicle-to-everything (v2x) (release 16),'' \emph{3GPP TR 36.885}, 2019.

\bibitem{wilabV2Xsim}
V.~Todisco, S.~Bartoletti, C.~Campolo, A.~Molinaro, A.~O. Berthet, and A.~Bazzi, ``Performance analysis of sidelink 5g-v2x mode 2 through an open-source simulator,'' \emph{IEEE Access}, vol.~9, pp. 145\,648--145\,661, 2021.

\bibitem{extendedDataYang}
J.~M. Yang, H.~Yoon, S.~Hwang, and S.~Bahk, ``Press: Predictive assessment of resource usage for c-v2v mode 4,'' in \emph{2021 IEEE Wireless Communications and Networking Conference (WCNC)}, 2021, pp. 1--6.

\bibitem{extendedDataZhang}
F.~Zhang, M.~M. Wang, and R.~Deng, ``On reliability bound and improvement of sensing-based semipersistent scheduling in lte-v2x,'' \emph{IEEE Internet of Things Journal}, vol.~8, no.~7, pp. 6101--6113, 2021.

\bibitem{extendedDataAli1}
M.~Ali, H.~Hwang, and Y.-T. Kim, ``Enhanced c-v2x mode-4 with virtual cell, resource usage bitmap, and smart roaming,'' \emph{IEEE Access}, vol.~11, pp. 142\,628--142\,642, 2023.

\bibitem{extendedDataAli2}
M.~Ali and Y.-T. Kim, ``Cognitive collision resolution for enhanced performance in c-v2x sidelink mode 4,'' in \emph{2021 22nd Asia-Pacific Network Operations and Management Symposium (APNOMS)}, 2021, pp. 102--107.

\bibitem{extendedDataAli3}
M.~Ali, H.~Hwang, and Y.-T. Kim, ``Performance enhancement of c-v2x mode 4 with balanced resource allocation,'' in \emph{ICC 2022 - IEEE International Conference on Communications}, 2022, pp. 2750--2755.

\bibitem{extendedDataSabeeh1}
S.~Sabeeh, P.~Sroka, and K.~Wesołowski, ``Estimation and reservation for autonomous resource selection in c-v2x mode 4,'' in \emph{2019 IEEE 30th Annual International Symposium on Personal, Indoor and Mobile Radio Communications (PIMRC)}, 2019, pp. 1--6.

\bibitem{extendedDataSabeeh2}
S.~Sabeeh and K.~Wesołowski, ``C-v2x mode 4 resource allocation in high mobility vehicle communication,'' in \emph{2020 IEEE 31st Annual International Symposium on Personal, Indoor and Mobile Radio Communications}, 2020, pp. 1--6.

\bibitem{extendedDataSabeeh3}
------, ``Congestion control in autonomous resource selection of cellular-v2x,'' \emph{IEEE Access}, vol.~11, pp. 7450--7460, 2023.

\bibitem{extendedDataWang1}
B.~Wang, J.~Zheng, N.~Mitton, and C.~Li, ``An enhanced c-v2x mode 4 resource selection scheme for cav platoons in a multilane highway scenario,'' in \emph{2023 International Conference on Future Communications and Networks (FCN)}, 2023, pp. 1--6.

\bibitem{extendedDataWang2}
------, ``Inp-crs: An intra-platoon cooperative resource selection scheme for c-v2x networks,'' \emph{IEEE Communications Letters}, vol.~27, no.~11, pp. 3118--3122, 2023.

\bibitem{selectProcedureBazzi}
A.~Bazzi, C.~Campolo, A.~Molinaro, A.~O. Berthet, B.~M. Masini, and A.~Zanella, ``On wireless blind spots in the c-v2x sidelink,'' \emph{IEEE Transactions on Vehicular Technology}, vol.~69, no.~8, pp. 9239--9243, 2020.

\bibitem{selectProcedureMolina}
R.~Molina-Masegosa, M.~Sepulcre, and J.~Gozalvez, ``Geo-based scheduling for c-v2x networks,'' \emph{IEEE Transactions on Vehicular Technology}, vol.~68, no.~9, pp. 8397--8407, 2019.

\bibitem{ccncYang}
W.~Yang, B.~Jeon, C.~Mun, and H.-S. Jo, ``Cellular-v2x qos adaptive distributed congestion control: A deep q network approach,'' in \emph{2023 IEEE 20th Consumer Communications \& Networking Conference (CCNC)}, 2023, pp. 967--968.

\bibitem{drlYang}
\BIBentryALTinterwordspacing
W.~Yang and H.-S. Jo, ``Deep-reinforcement-learning-based range-adaptive distributed power control for cellular-v2x,'' \emph{ICT Express}, vol.~9, no.~4, pp. 648--655, 2023. [Online]. Available: \url{https://www.sciencedirect.com/science/article/pii/S2405959522001059}
\BIBentrySTDinterwordspacing

\bibitem{jychoi}
J.-Y. Choi, H.-S. Jo, C.~Mun, and J.-G. Yook, ``Deep reinforcement learning-based distributed congestion control in cellular v2x networks,'' \emph{IEEE Wireless Communications Letters}, vol.~10, no.~11, pp. 2582--2586, 2021.

\bibitem{aiBasedSelectSang}
J.~Sang, T.~Zhou, T.~Xu, Y.~Jin, and Z.~Zhu, ``Deep learning based predictive power allocation for v2x communication,'' \emph{IEEE Access}, vol.~9, pp. 72\,881--72\,893, 2021.

\bibitem{aiBasedSelectReshma}
\BIBentryALTinterwordspacing
P.~Reshma and V.~Sudha, ``Cell edge throughput enhancement in v2x communications using graph-based advanced deep learning scheduling algorithms,'' \emph{IETE Journal of Research}, vol.~0, no.~0, pp. 1--16, 2024. [Online]. Available: \url{https://doi.org/10.1080/03772063.2024.2351544}
\BIBentrySTDinterwordspacing

\bibitem{drlBasedSelectLiang}
L.~Liang, H.~Ye, and G.~Y. Li, ``Spectrum sharing in vehicular networks based on multi-agent reinforcement learning,'' \emph{IEEE Journal on Selected Areas in Communications}, vol.~37, no.~10, pp. 2282--2292, 2019.

\bibitem{drlBasedSelectJu}
Y.~Ju, Y.~Chen, Z.~Cao, L.~Liu, Q.~Pei, M.~Xiao, K.~Ota, M.~Dong, and V.~C.~M. Leung, ``Joint secure offloading and resource allocation for vehicular edge computing network: A multi-agent deep reinforcement learning approach,'' \emph{IEEE Transactions on Intelligent Transportation Systems}, vol.~24, no.~5, pp. 5555--5569, 2023.

\bibitem{drlBasedSelectLee}
I.~Lee and D.~K. Kim, ``Decentralized multi-agent dqn-based resource allocation for heterogeneous traffic in v2x communications,'' \emph{IEEE Access}, vol.~12, pp. 3070--3084, 2024.

\bibitem{drlBasedSelectFarzanullah}
M.~Farzanullah and T.~Le-Ngoc, ``Platoon leader selection, user association and resource allocation on a c-v2x based highway: A reinforcement learning approach,'' in \emph{ICC 2023 - IEEE International Conference on Communications}, 2023, pp. 5396--5401.

\bibitem{physicsBasedAI}
\BIBentryALTinterwordspacing
N.~Thuerey, P.~Holl, M.~Mueller, P.~Schnell, F.~Trost, and K.~Um, \emph{Physics-based Deep Learning}.\hskip 1em plus 0.5em minus 0.4em\relax WWW, 2021. [Online]. Available: \url{https://physicsbaseddeeplearning.org}
\BIBentrySTDinterwordspacing

\bibitem{tsnamTIFS}
T.~Nam, D.-H. Choi, E.~Lee, H.-S. Jo, and J.-G. Yook, ``Data generation and augmentation method for deep learning-based vdu leakage signal restoration algorithm,'' \emph{IEEE Transactions on Information Forensics and Security}, vol.~19, pp. 5220--5234, 2024.

\bibitem{SUMO}
\BIBentryALTinterwordspacing
P.~A. Lopez, M.~Behrisch, L.~Bieker-Walz, J.~Erdmann, Y.-P. Fl{\"o}tter{\"o}d, R.~Hilbrich, L.~L{\"u}cken, J.~Rummel, P.~Wagner, and E.~Wie{\ss}ner, ``Microscopic traffic simulation using sumo,'' in \emph{The 21st IEEE International Conference on Intelligent Transportation Systems}.\hskip 1em plus 0.5em minus 0.4em\relax IEEE, 2018. [Online]. Available: \url{https://elib.dlr.de/124092/}
\BIBentrySTDinterwordspacing

\bibitem{accRef1}
\BIBentryALTinterwordspacing
S.~E. Shladover, ``{Longitudinal Control of Automated Guideway Transit Vehicles Within Platoons},'' \emph{Journal of Dynamic Systems, Measurement, and Control}, vol. 100, no.~4, pp. 302--310, 12 1978. [Online]. Available: \url{https://doi.org/10.1115/1.3426382}
\BIBentrySTDinterwordspacing

\bibitem{accRef2}
R.~Rajamani and C.~Zhu, ``Semi-autonomous adaptive cruise control systems,'' \emph{IEEE Transactions on Vehicular Technology}, vol.~51, no.~5, pp. 1186--1192, 2002.

\bibitem{accRef3}
\BIBentryALTinterwordspacing
G.~Naus, J.~Ploeg, M.~{Van de Molengraft}, W.~Heemels, and M.~Steinbuch, ``Design and implementation of parameterized adaptive cruise control: An explicit model predictive control approach,'' \emph{Control Engineering Practice}, vol.~18, no.~8, pp. 882--892, 2010. [Online]. Available: \url{https://www.sciencedirect.com/science/article/pii/S0967066110000882}
\BIBentrySTDinterwordspacing

\bibitem{accRef4}
F.~A. Mullakkal-Babu, M.~Wang, B.~van Arem, and R.~Happee, ``Design and analysis of full range adaptive cruise control with integrated collision a voidance strategy,'' in \emph{2016 IEEE 19th International Conference on Intelligent Transportation Systems (ITSC)}, 2016, pp. 308--315.

\bibitem{caccETSI}
``Intelligent transport system (its); cooperative adaptive cruise control (cacc); pre-standardization study,'' Tech. Rep., 2019.

\bibitem{caccTR}
T.~Tiernan, N.~Richardson, P.~Azeredo, W.~G. Najm, T.~Lochrane \emph{et~al.}, ``Test and evaluation of vehicle platooning proof-of-concept based on cooperative adaptive cruise control,'' John A. Volpe National Transportation Systems Center (US), Tech. Rep., 2017.

\bibitem{platoonTR}
``Study enhancement 3gpp support for 5g v2x services,'' John A. Volpe National Transportation Systems Center (US), Tech. Rep., 2018.

\bibitem{caccPractical}
H.~Xing, B.~Cimoli, I.~Passchier, G.~Kakes, V.~Ho, and H.~Nijmeijer, ``Practical challenges in cacc communication: Its g5, lte uu, and lte sidelink pc5,'' in \emph{2021 European Control Conference (ECC)}, 2021, pp. 1795--1801.

\bibitem{caccTITS2}
M.~Sybis, V.~Vukadinovic, M.~Rodziewicz, P.~Sroka, A.~Langowski, K.~Lenarska, and K.~Wesołowski, ``Communication aspects of a modified cooperative adaptive cruise control algorithm,'' \emph{IEEE Transactions on Intelligent Transportation Systems}, vol.~20, no.~12, pp. 4513--4523, 2019.

\bibitem{ccNEDO}
M.~Omae, R.~Fukuda, T.~Ogitsu, and W.-P. Chiang, ``Control procedures and exchanged information for cooperative adaptive cruise control of heavy-duty vehicles using broadcast inter-vehicle communication,'' \emph{International journal of intelligent transportation systems research}, vol.~12, pp. 84--97, 2014.

\bibitem{caccTITS3}
Y.~Zhang, Y.~Bai, M.~Wang, and J.~Hu, ``Cooperative adaptive cruise control with robustness against communication delay: An approach in the space domain,'' \emph{IEEE Transactions on Intelligent Transportation Systems}, vol.~22, no.~9, pp. 5496--5507, 2021.

\bibitem{caccTITS4}
L.~Cui, Z.~Chen, A.~Wang, J.~Hu, and B.~B. Park, ``Development of a robust cooperative adaptive cruise control with dynamic topology,'' \emph{IEEE Transactions on Intelligent Transportation Systems}, vol.~23, no.~5, pp. 4279--4290, 2022.

\bibitem{joiningLeaving}
A.~Farag, D.~M. Mahfouz, O.~M. Shehata, and E.~I. Morgan, ``A novel ros-based joining and leaving protocols for platoon management,'' in \emph{2019 IEEE International Conference on Vehicular Electronics and Safety (ICVES)}, 2019, pp. 1--6.

\bibitem{sbsps}
A.~Nabil, K.~Kaur, C.~Dietrich, and V.~Marojevic, ``Performance analysis of sensing-based semi-persistent scheduling in c-v2x networks,'' in \emph{2018 IEEE 88th Vehicular Technology Conference (VTC-Fall)}, 2018, pp. 1--5.

\bibitem{camRef1}
\BIBentryALTinterwordspacing
J.~Santa, F.~Pereñíguez, A.~Moragón, and A.~F. Skarmeta, ``Experimental evaluation of cam and denm messaging services in vehicular communications,'' \emph{Transportation Research Part C: Emerging Technologies}, vol.~46, pp. 98--120, 2014. [Online]. Available: \url{https://www.sciencedirect.com/science/article/pii/S0968090X14001193}
\BIBentrySTDinterwordspacing

\bibitem{camRef2}
``ntelligent transport systems (its); vehicular communications; basic set of applications; part 2: Specification of cooperative awareness basic service,'' Tech. Rep., 2011.

\bibitem{camRef3}
V.~de~C{\'o}zar, J.~Poncela, M.~Aguilera, M.~Aamir, and B.~S. Chowdhry, ``Cooperative vehicle-to-vehicle awareness messages implementation,'' in \emph{Wireless Sensor Networks for Developing Countries}, F.~K. Shaikh, B.~S. Chowdhry, H.~M. Ammari, M.~A. Uqaili, and A.~Shah, Eds.\hskip 1em plus 0.5em minus 0.4em\relax Berlin, Heidelberg: Springer Berlin Heidelberg, 2013, pp. 26--37.

\bibitem{tsnamVTC}
T.~Nam, S.~Lee, N.~Jeong, H.-S. Jo, and J.-G. Yook, ``Proactive resource allocation in c-v2x for cacc-based platoon driving service,'' in \emph{2024 IEEE 100th Vehicular Technology Conference (VTC2024-Fall)}, 2024, pp. 1--5.

\end{thebibliography}

\begin{IEEEbiography}[{\includegraphics[width=1in,height=1.25in,clip,keepaspectratio]{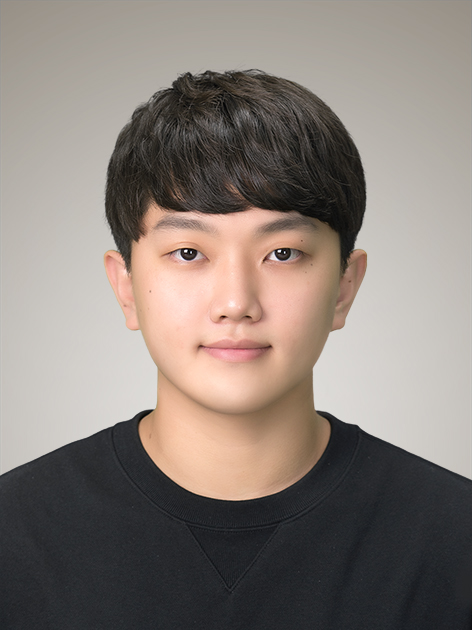}}]{Taesik Nam (Student Member, IEEE)} 
received the B.S. degree in electronics and control engineering from Hanbat National University, Daejeon, South Korea, in 2020. He is currently pursuing the Ph.D. degree in electrical and electronic engineering with Yonsei University, Seoul, South Korea. His research interests include the machine learning, information recovery, signal processing, wireless communication, and autonomous vehicle.
\end{IEEEbiography}

\begin{IEEEbiography}[{\includegraphics[width=1in,height=1.25in,clip,keepaspectratio]{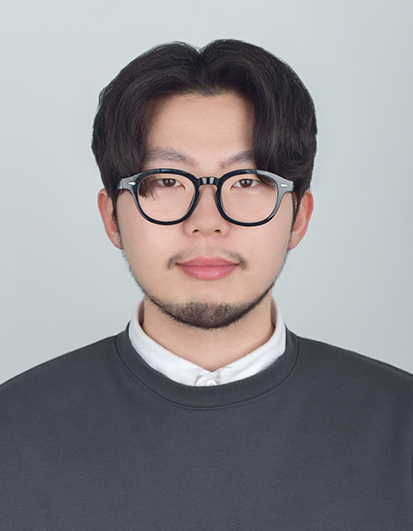}}]{Seungjae Lee} 
received the B.S. degree in electronics and control engineering from Hanbat National University, Daejeon, South Korea. He is currently pursuing the M.S. degree in automotive engineering with Hanyang University, Seoul, South Korea. His research interests include machine learning, wireless communication, and autonomous vehicles.
\end{IEEEbiography}

\begin{IEEEbiography}[{\includegraphics[width=1in,height=1.25in,clip,keepaspectratio]{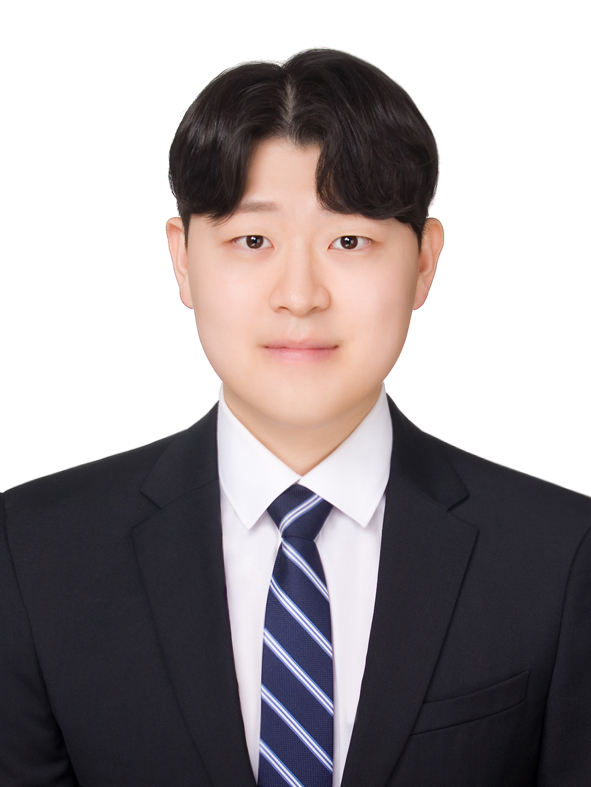}}]{Kiwoong Park} 
received the B.S. degree in electronics and control engineering from Hanbat National University, Daejeon, South Korea., in 2024. He is currently pursuing the M.S. degree in automotive engineering with Hanyang University, Seoul, South Korea. His research interests include machine learning, wireless communication, and autonomous vehicles.
\end{IEEEbiography}

\begin{IEEEbiography}[{\includegraphics[width=1in,height=1.25in,clip,keepaspectratio]{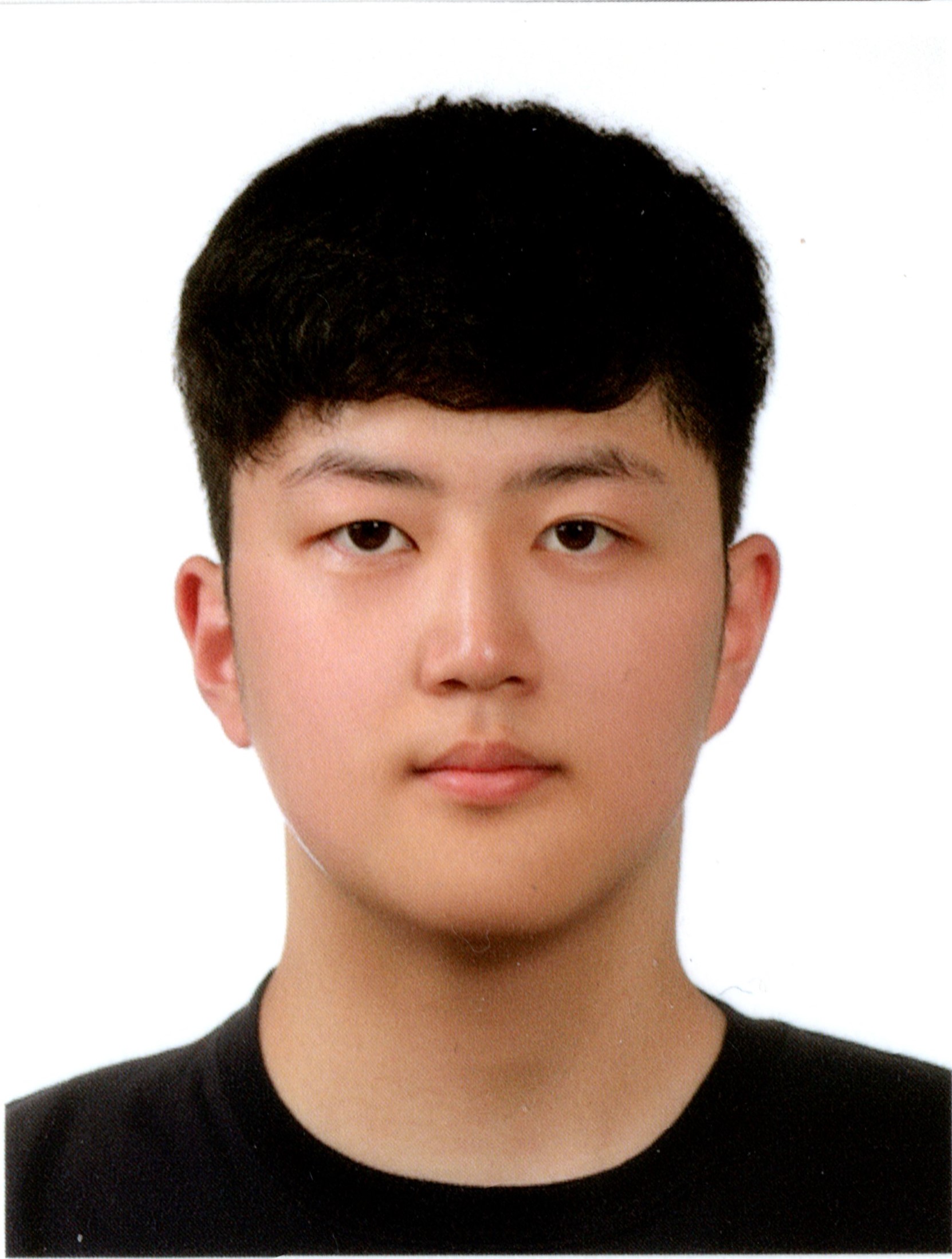}}]{Sunbeom Kwon} 
received the B.S degree in Department of Information and Communication Engineering at Chungbuk National University(CBNU) in Cheongju, South Korea in 2024. He is currently pursuing a M.S degree in the Department of Automotive Engineering at Hanyang University, Seoul,South Korea. His research interests include autonomous vehicle simulators and V2X communication.
\end{IEEEbiography}

\begin{IEEEbiography}
[{\includegraphics[width=1in,height=1.25in,clip,keepaspectratio]{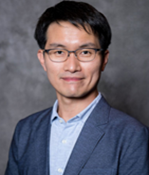}}]{Nathan Jeong}
(M’07-SM’18) received the B.S. degree from Korea Maritime University, Busan, South Korea, in 2000, the M.S. degree from Yonsei University, Seoul, South Korea, in 2002, and the Ph.D. degree from Purdue University, West Lafayette, IN, USA, in 2010. He was a visiting scholar in Georgia Institute of Technology, Atlanta, GA, USA, in 2010.  He is currently an assistant professor at The University of Alabama. His current research interests include electromagnetic device and system, remote sensing radar, microwave and millimeter-wave imaging, wireless power transmission, antenna, vehicle-to-everything (V2X) communication, and wearable sensors for biomedical application. He has total of 12 years of industrial experience at Samsung Electronics, BlackBerry, and Qualcomm. He has authored or co-authored over 49 technical publications in in refereed journals and conference proceedings. In addition, he holds 67 international patent and patent applications in the areas of antennas, wireless communication circuit, microwave and millimeter-wave system, V2X, wireless power transfer, and bioelectronics. He received the fifth place at Qualcomm Research Paper Competition in 2015, the Honorable Mention Award in RF Alliance in 2010, the Honorable Mention Paper Award from IEEE AP-S in 2008, and Samsung Humantech Paper Award in 2002.
\end{IEEEbiography}

\begin{IEEEbiography}[{\includegraphics[width=1in,height=1.25in,clip,keepaspectratio]{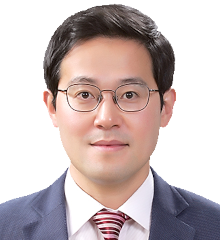}}]{Han-Shin Jo}
(S’06-M’10) received B.S., M.S., and Ph.D. degrees in electrical and electronics engineering from Yonsei University, Seoul, South Korea, in 2001, 2004, and 2009, respectively. He is currently a Professor at the Department of Automotive Engineering, Hanyang University, Seoul, Republic of Korea. He was a professor at the Department of Electronic Engineering, Hanbat National University, Daejeon, Republic of Korea from 2012 to 2023. He was a Post-Doctoral Research Fellow with the Wireless Networking and Communications Group, Department of Electrical and Computer Engineering, University of Texas at Austin, from 2009 to 2011. He developed a long-term evolution base station in Samsung Electronics from 2011 to 2012. He has been a member of the Korea ITU-R Working Party 5D Committee since 2016. He was a visiting scholar with the Wireless Devices and Systems Group at the University of Southern California, from 2018 to 2019. His current research interests are in coexistence study and spectrum sharing, and applications of stochastic geometry, optimization theory, and machine/reinforcement learning to wireless communications and connected and autonomous mobility.
\end{IEEEbiography}

\begin{IEEEbiography}[{\includegraphics[width=1in,height=1.25in,clip,keepaspectratio]{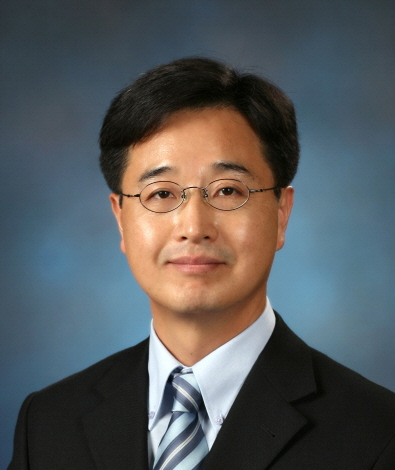}}]{Jong-gwan Yook} (S’89–M’97–SM’12) received the B.S. and M.S. degrees in electronics engineering from Yonsei University, Seoul, Korea, and the Ph.D. degree in electrical engineering and computer science from the University of Michigan, Ann Arbor, MI, USA, in 1987, 1989, and 1996, respectively. He is currently a Professor with the School of Electrical and Electronic Engineering, Yonsei University. His research interests include theoretical/numerical EM modeling and characterization of microwave/millimeter-wave circuits and components and design, analysis, and optimization of high-frequency high-speed interconnects, including signal/power integrity (EMI/EMC), based on frequency-domain as well as time-domain full-wave methods. Prof. Yook was the recipient of the Excellent Teaching and Research Activity Award from Yonsei University several times. From 2009 to 2012, he was the Chair of the Korean EMC Society. From 2012 to 2013, he was a Distinguished Lecturer for the IEEE EMC Society. He was also the Chair of the Technical Program Committee of Asia Pacific International Symposium on Electromagnetic Compatibility Conference 2017. In 2023, he served as the President of the Korean Institute of Electromagnetic Engineering and Science (KIEES).
\end{IEEEbiography}
\end{document}